\documentclass[aps,prd,preprint,showpacs,floatfix,preprintnumbers,nofootinbib,superscriptaddress,
showkeys]{revtex4}
\usepackage[utf8]{inputenc}
\usepackage{fancybox}
\usepackage{hhline}
\usepackage{dcolumn}
\usepackage{textcomp}
\usepackage{epsfig,graphics,graphicx}
\usepackage{amsfonts,amssymb,amsmath}
\usepackage{pifont}
\usepackage{bm}
\usepackage{longtable} 
\usepackage{appendix}
\usepackage{lscape}
\usepackage[mathscr]{euscript}
\usepackage{mathrsfs}
\usepackage{multirow}
\usepackage{rotating}
\usepackage{color}

\newcommand{\beq}{\begin{equation}}
\newcommand{\eeq}{\end{equation}}
\newcommand{\bea}{\begin{eqnarray}}
\newcommand{\eea}{\end{eqnarray}}

\newcommand{\eps}{\epsilon}
\newcommand{\ord}[1]{{\cal{O}}( #1 )}
\newcommand{\B}{{\bf B}}
\newcommand{\Bdag}{{\bf B^\dagger}}
\newcommand{\p}{{\mathfrak{p}}_0}

\DeclareFontFamily{OT1}{pzc}{}
\DeclareFontShape{OT1}{pzc}{m}{it}%
              {<-> s * [0.900] pzcmi7t}{}
\DeclareMathAlphabet{\mathpzc}{OT1}{pzc}%
                                 {m}{it}
\DeclareMathAlphabet{\mathcalligra}{T1}{calligra}{m}{n}
\usepackage{hyperref}
\hypersetup{pdfauthor={me}, colorlinks=true, citecolor=blue, urlcolor=blue, linkcolor=black}
\begin{document}
\preprint{\vbox{\hbox{ JLAB-THY-18-3} }}
\title{\phantom{x}
\vspace{-0.5cm}     }
\title{  Baryon  Chiral Perturbation Theory    combined with the ${\mathbf{1/N_c}}$ Expansion in SU(3) I:  Framework   }
\author{I.~P.~Fernando}\email{ishara@jlab.org }
\author{J.~L.~Goity}\email{goity@jlab.org}
\affiliation{Department of Physics, Hampton University, Hampton, VA 23668, USA. }
\affiliation{Thomas Jefferson National Accelerator Facility, Newport News, VA 23606, USA.}
%\date{\today}
 \begin{abstract}
Baryon Chiral Perturbation Theory  combined with the  $1/N_c$ expansion  is implemented for   three flavors.  Baryon masses, vector charges and axial vector couplings are studied to one-loop and organized according to   the $\xi$-expansion, in which  the $1/N_c$ and the  low energy power countings are linked according to $1/N_c=\ord{\xi}=\ord{p}$.  The   renormalization to $\ord{\xi^3}$ necessary for the mentioned observables is provided,  along with applications to the baryon masses and axial couplings as obtained in lattice QCD calculations.

\end{abstract}


\maketitle

\tableofcontents

\section{Introduction}
\label{sec:Intro}

The low energy effective theory for baryons is a recurrent topic in low energy QCD,  which  has evolved   through different approaches and improvements. The original version of baryon  Chiral Perturbation Theory (ChPT)
\cite{Pagels:1974se}
 gave rise to different versions of baryon effective field theories based on effective chiral Lagrangians \cite{Weinberg:1968de,Coleman:1969sm,Callan:1969sn}, starting with the relativistic version 
\cite{Gasser:1987rb,Krause:1990xc} 
or   Baryon ChPT (BChPT), followed by the non-relativistic version based in an expansion in the inverse baryon mass 
\cite{Jenkins:1990jv,Bernard:1992qa,Bernard:1995dp,Ecker:1995rk}
 or   Heavy Baryon ChPT (HBChPT), and  by  manifestly Lorentz covariant versions based on the IR regularization scheme \cite{Ellis:1997kc,Becher:1999he,Fuchs:2003qc}, which allow for an explicit implementation of the low energy power counting. 
  In all those versions of the baryon effective theory a consistent low energy expansion can be implemented.  A key issue,  which became apparent quite early,  was the convergence of the low energy expansion. Being an expansion that progresses in steps of $\ord{p}$,  in contrast to the expansion in the pure Goldstone Boson sector where the steps are $\ord{p^2}$, it is natural to expect a slower rate of convergence. However, a key factor  affecting  the convergence   has to do with the  relatively small mass gap  between the spin 1/2 and 3/2 baryons. In the context of BChPT, it was realized in \cite{Jenkins:1991es} that the inclusion of the spin 3/2  degrees of freedom   improves the convergence of the one-loop contributions to certain observables such as  the $\pi$-$N$ scattering amplitude and the axial currents and magnetic moments.  There have been since then numerous works including spin 3/2 baryons \cite{Hemmert:1996xg,Hemmert:1997ye,Hemmert:2003cb,Fettes:2000bb,Procura:2006bj,Hacker:2005fh,Bernard:2003xf,Bernard:2005fy,Procura:2006gq,Pascalutsa:2007yg}.  The explanation of  those improvements was obtained through   the study of baryons in the large $N_c$ limit of QCD \cite{'tHooft:1973jz}, where in that limit a dynamical spin-flavor symmetry  emerges  \cite{Gervais:1983wq,Gervais:1984rc,Dashen:1993as,Dashen:1993ac}, which  requires the inclusion of the higher spin baryons in the effective theory and  leads to a better behaved low energy expansion. In the large $N_c$ limit, baryons behave very differently than mesons \cite{Witten:1979kh}, in particular because their masses scale like $\ord{N_c}$ (they are the heavy sector of QCD) and the $\pi$-baryon couplings are $\ord{\sqrt{N_c}}$. Those properties were shown to demand,  for consistency with $\pi$-baryon scattering at large $N_c$, that at large $N_c$ baryons must respect the mentioned dynamical contracted spin-flavor symmetry $SU(2 N_f)$, $N_f$ being the number of light flavors \cite{Gervais:1983wq,Gervais:1984rc,Dashen:1993as,Dashen:1993ac},  which is broken by effects ordered in powers of $1/N_c$ and in  powers of the quark mass differences. The inclusion of the consistency requirements of the large $N_c$ limit into the effective theory came naturally through a combination of the $1/N_c$ expansion and  HBChPT 
\cite{Jenkins:1995gc},  which is the framework followed in the present work. The study of one-loop corrections in that framework 
 was first  carried out in Refs.~\cite{Jenkins:1995gc,FloresMendieta:1998ii,FloresMendieta:2006ei} and more recently in  \cite{CalleCordon:2012xz,Cordon:2013era}.  
   In the  combined theory   the $1/N_c$ and Chiral expansions do not commute \cite{Cohen:1992uy}: the reason is  the baryon  mass splitting scale of $\ord{1/N_c}$ ($\Delta-N$ mass difference), for which it becomes necessary to specify its order in terms of the low energy expansion. Thus the $1/N_c$ and Chiral expansions must be linked. Particular emphasis will be given to the specific linking in which the baryon mass splitting is taken to be $\ord{p}$ in the Chiral expansion, and which will be called the $\xi$-expansion. 
     Following references   \cite{Jenkins:1995gc,FloresMendieta:1998ii,FloresMendieta:2006ei,CalleCordon:2012xz}, in the present work  the framework for HBChPT$\times 1/N_c$ is extended to three flavors. The renormalization necessary for the baryon masses, and the vector charges and axial-vector 
currents is implemented to    one-loop, i.e., $\ord{\xi^3}$.  As it had been done in the case of two flavors \cite{CalleCordon:2012xz}, the present work  gives all results at generic values of $N_c$, i.e., all formulas presented have been derived for general $N_c$, and therefore  detailed analyses of  $N_c$ dependencies can be carried out.

The  significant   progress in  lattice QCD   (LQCD) calculations of baryon observables  \cite{Hagler:2009ni,Fodor:2012gf,Alexandrou:2011iu} provides opportunities for further testing and understanding  low energy effective theories of baryons, which in turn can serve to understand the LQCD results themselves. The   determination of  the quark mass dependence of the various low energy observables, such as masses, axial couplings, magnetic moments, electromagnetic polarizabilities, etc.,  are of key importance for   testing    the effective theory, in particular  its range of validity in quark masses,  as well as for the determination of its low energy constants (LECs). Lattice results for  $N$ and $\Delta$ as well as hyperon masses  \cite{Durr:2008zz, WalkerLoud:2008bp, Aoki:2008sm, Lin:2008pr, Alexandrou:2009qu, Aoki:2009ix, Aoki:2010dy, Bietenholz:2011qq, Alexandrou:2014sha} (results of the last reference are used in the present work),  the axial coupling $g_A$ of the nucleon \cite{Edwards:2005ym,Bratt:2010jn,Alexandrou:2010hf,Yamazaki:2008py,Yamazaki:2009zq,Lin:2008uz}  and a subset of  the axial couplings of the octet and decuplet baryons \cite{Alexandrou:2016xok} at varying quark masses can be  analyzed with     the effective theory, as presented in this work.
 
This work is organized as follows. In Section~\ref{sec:Framework} the framework for the combined   ${ { 1/N_c}}$ and HBChPT expansions is described.
Section~\ref{sec:Masses} presents the evaluation  of the baryon masses to $\ord{\xi^3}$, Section~\ref{sec:Vector} presents the corrections to the vector charges, and Section~\ref{sec:Axial} the corrections to the  axial couplings. In both Sections~\ref{sec:Masses}  and \ref{sec:Axial} applications to  LQCD results are presented. Finally,  a summary is given in Section~\ref{sec:Conclusions}. Several appendices present useful material needed in the calculations,  namely,  
Appendix~\ref{app:Algebra} on spin-flavor algebra, 
Appendix~\ref{app:buildblocks} on tools to build the chiral Lagrangians, 
Appendix~\ref{app:loopintegrals} on the one-loop integrals, and 
Appendix~\ref{app:opred} on   reduction formulas of composite operators.

\section{Combined Baryon Chiral Perturbation Theory and ${\mathbf{\rm 1/N_c}}$ expansion for three flavors }
\label{sec:Framework}

In this section the framework for the combined $1/N_c$ and chiral expansions in baryons is presented in some detail along similar lines as in the original works \cite{Jenkins:1995gc,FloresMendieta:1998ii,FloresMendieta:2006ei} and the more recent work \cite{CalleCordon:2012xz,Cordon:2013era}. The symmetries that constrain the effective Lagrangian in the   chiral    and large $N_c$ limits are chiral $SU_L(N_f)\times SU_R(N_f)$, which is a Noether symmetry, and contracted dynamical spin-flavor symmetry $SU(2N_f)$\cite{Gervais:1983wq,Gervais:1984rc,Dashen:1993ac,Dashen:1993as}~\footnote{See also Appendix~\ref{app:Algebra}.}. $N_f$ is the number of light flavors, where in this work $N_f=3$.  In the limit $N_c\to \infty$ the spin-flavor symmetry requires baryon states to fill degenerate multiplets of $SU(6)$. In particular, the  ground state (GS)  baryons belong into a symmetric $SU(6)$ multiplet.
At finite $N_c$ the spin-flavor symmetry is broken by effects suppressed by powers of $1/N_c$,  and the   mass splittings  in the GS multiplet between the states with spins $S+1$ and   $S$ are proportional to $(S+1)/N_c$. The effects of finite $N_c$ are then implemented as an expansion in $1/N_c$  in  the effective Lagrangian.  Because baryon masses are proportional to $N_c$, it becomes natural to use the framework of HBChPT \cite{Jenkins:1990jv,Jenkins:1991ne}, where the expansion in inverse powers of the baryon mass becomes part of the $1/N_c$ expansion. The framework used here  follows that of Refs. \cite{Jenkins:1995gc,FloresMendieta:1998ii,CalleCordon:2012xz}.

The dynamical contracted $SU(2N_f)$ symmetry results from the requirement of large $N_c$ consistency of baryon observables \cite{Gervais:1983wq,Gervais:1984rc,Dashen:1993ac,Dashen:1993as}~\footnote{See also Appendix~\ref{app:Algebra}.}, in particular the requirement that the Born contribution to the Goldstone Boson-baryon (GB-baryon) scattering amplitude be finite as $N_c\to \infty$. The constraint emerges because the GB-baryon coupling is $\ord{\sqrt{N_c}}$, and therefore cancellations between crossed diagrams must occur. The 35 generators of $SU(6)$ and their commutation relations are the following:
\bea
&& S^i: \text{ SU(2) spin generators, ~~} T^a: \text{ SU(3) flavor generators,~~} G^{ia}: \text{ spin-flavor generators}\nonumber\\
&&[S^i,S^j]=i\eps^{ijk} S^k\nonumber\\
&&[T^a,T^b]=if^{abc}T^c\nonumber\\
&&[S^i,T^a]=0,~~~~[S^i,G^{ja}]=i \eps^{ijk}G^{ka},~~~~[T^a,G^{ib}]=if^{abc}G^{ic}\nonumber\\
&&[G^{ia},G^{jb}]=\frac{i}{4}\delta^{ij}f^{abc}T^c+\frac{i}{6}\delta^{ab}\eps^{ijk} S^k+\frac{i}{2}\eps^{ijk}d^{abc}G^{kc}.
\eea
The generators $G^{ia}$ have coherent matrix elements, i.e., matrix elements that scale as $N_c$ between baryons of spin $S=\ord{N_c^0}$. These generators are the ones that represent the spatial components of axial-vector currents at the leading order in the $1/N_c$ expansion.
A contracted $SU(6)$ symmetry, which is the actual dynamical symmetry in large $N_c$, is generated by the Algebra where   $G^{ia}$ is replaced by $X^{ia}\equiv G^{ia}/N_c$.
The ground state baryons belong to the totally symmetric spin-flavor irreducible representation with $N_c$ spin-flavor indices, and consist of states with spin $S=1/2,\cdots,N_c/2$ (assuming $N_c$ to be odd). For a given spin $S$ the corresponding $SU(3)$ multiplet is $(p,q)=(2S,\frac{1}{2} (N_c-2S))$ in the usual Young tableu notation. For $N_c=3$ the states are the physical $S=1/2$ octet and $S=3/2$ decuplet. 

In HBChPT the   baryon field, denoted by $\bf B$,  represents  the   spin-flavor multiplet where its components are sorted out by spin and flavor, that is, the entries in $\B$ have well defined spin, and therefore they are in irreducible representations of $SU(3)$. 

Implementing  chiral symmetry follows the well known scheme of  the non-linear realization on the matter fields. Representing the Goldstone Boson octet by:
\beq
u=e^{i \pi^a T^a/F_\pi},
\eeq
  the non-linear transformation law is implemented:
\beq
R \,u \,h^\dag(L,R,u)=h(L,R,u) \,u\, L^\dag,
\eeq
where $L$ ($R$) is a transformation of $SU_L(3)$ ( $SU_R(3)$ ). $h(L,R,u)$ is then a $SU(3)$ flavor transformation.
One can therefore define the usual chiral transformations on the baryon fields according to:
\beq
(L,R):{\bf B}=h(L,R,u){\bf B},
\label{eq:LR-transf}
\eeq
where obviously the non-linear transformation $h$ acts on the different components of $\bf B$ with the corresponding $SU(3)$ irreducible representation. Chiral transformations do not commute with $SU(6)$, but they leave the commutation relations unchanged.
The chiral covariant derivative $D_\mu {\bf B}$ is then given by:
\bea
D_\mu \B&=&\partial_\mu \B-i \Gamma_\mu  \B ,\nonumber\\
\Gamma_\mu&=&\frac{1}{2}\,(u^\dag(i\partial_\mu+r_\mu) u+u(i\partial_\mu+ l_\mu) u^\dag),
\label{eq:Dmu}
\eea
where   $l_\mu  = v_\mu - a_\mu$  and $r_\mu = v_\mu + a_\mu$   are gauge sources.  
Another     building block   is the axial Maurer-Cartan one-form:
\beq
u_\mu=u^\dag(i\partial_\mu+r_\mu) u-u(i\partial_\mu+ l_\mu) u^\dag,~~~~(L,R):u_\mu=h(L,R,u)u_\mu h^\dagger(L,R,u).
\label{eq:Maurer-Cartan}
\eeq
Both $\Gamma_\mu$ and $u_\mu$ belong to the $SU(3)$ Algebra, and are written in the general form
$X=X^a T^a$. When acting on the different components of the field $\B$, $T^a$ is obviously taken in the corresponding $SU(3)$ irreducible representation.

The scalar and pseudoscalar   densities are collected into:
\bea
\chi&=&2 B_0(s+i p)\nonumber\\
\chi_{\pm}&\equiv&u^\dag  \chi u^\dag \pm u\chi^\dag  u\nonumber\\
\chi_\pm^0&=&\langle \chi_\pm\rangle\nonumber\\
\tilde \chi_\pm&\equiv& \chi_\pm^a T^a, 
\eea
where 
$s$ and $p$ are the scalar and pseudoscalar sources, and eventually $s$ is set to be the quark mass matrix.

The field strengths associated with the gauge sources are:
\bea
F^{\mu\nu}_{L}&=&\partial^\mu\ell^\nu-\partial^\nu\ell^\mu-i[\ell^\mu,\ell^\nu],~~F^{\mu\nu}_{R}=\partial^\mu r^\nu-\partial^\nu r^\mu-i[r^\mu,r^\nu]\nonumber\\
F^{\mu\nu}_{ \pm}&=&u F^{\mu\nu}_{L} u^\dag \pm u^\dag {F^{\mu\nu}_{R}}^\dag  u .  
\eea

Since contracted $SU(6)$ is not a Noether symmetry, its role in the effective Lagrangian is to primarily constrain couplings. For instance, at the leading order one such a constraint is that the GB-baryon  couplings   are determined by a single coupling $\mathring g_A$. The effective Lagrangian will be explicitly  invariant under rotations and chiral transformations and the QCD discrete symmetries $P$ and $T$.   
The Lagrangian consists of terms which are the product of tensors containing the GB  and source fields (chiral tensor operators) with terms which are   composite spin-flavor tensor operators built with products of   $SU(6)$ generators.
The $N_c$ power assigned to a term in the Lagrangian is determined by the spin-flavor operator according to $N_c^{1-n}$, where $n$ is the number of factors of $SU(6)$ generators involved in the operator. In general the chiral tensor operators carry hidden $N_c$ dependencies through the factors of $1/F_\pi$ accompanying the GB field operators, where $F_\pi=\ord{\sqrt{N_c}}$. Matrix elements of the spin-flavor operators carry additional $N_c$ dependencies, as is the case of operators where factors of the generators $G^{ia}$ appear, which lead to additional factors of $N_c$ in the matrix elements. Following this approach, the  Lagrangian terms are organized  in powers of the   chiral and $1/N_c$ expansions.  
The $1/N_c$ expansion naturally leads to the HBChPT expansion, as the large mass of the expansion is taken to be the spin-flavor singlet component of the baryon masses, namely $M_0=N_c\, m_0$ ($m_0$ can be considered here to be a LEC defined in the chiral limit and which will have itself an expansion in $1/N_c$). 

Bases of spin-flavor tensor operators are built using  the tools  in Appendix \ref{app:Algebra}, and requires in general   lengthy algebraic work. In the Appendix only the bases needed in this work are provided. 

In order to ensure the validity of the OZI rule for the quark mass dependency of baryon masses, namely, that the non-strange baryon mass dependence on $m_s$ is $\ord{N_c^0}$, the following combination of the source $\chi_+$      is defined:
\beq
\hat\chi_+\equiv \tilde\chi_+ +N_c\; \chi_+^0,
\eeq
which is $\ord{N_c}$ but has dependence on $m_s$ which is $\ord{N_c^0}$  for al states where the strangeness is $\ord{N_c^0}$.

For convenience a scale $\Lambda$ is introduced, which  can be chosen to be a typical  QCD scale, in order to render most of the LECs dimensionless. In the calculations $\Lambda=m_\rho$ will be chosen.

The lowest order Lagrangian  is  \cite{Jenkins:1995gc}:
 \bea
  {\cal{L}}_\B^{(1)}&=&\Bdag\left(i D_0 +  \mathring{g}_A
 u^{ia}G^{ia}-\frac{C_{\rm HF}}{N_c}{\hat {{S}}^2}+\frac{c_1}{2 \Lambda}  \;   \hat\chi_+ \right)\B.
\label{eq:Lagrangian-LO}
 \eea
 The kinetic term is $\ord{p\,N_c^0}$, and the terms involving GBs (when the vector and axial vector sources are turned off) start with the Weinberg-Tomozawa term which is $\ord{p/N_c}$. The second term gives in particular the axial vector current and the GB-baryon interaction.   $\mathring{g}_A$ is the axial coupling in the chiral and large $N_c$ limits (it has to be rescaled by a factor 5/6 to coincide with the usual axial coupling as defined for the nucleon, i.e., $g_A^N=g_A=\frac 56 \mathring{g}_A$). Because the matrix elements of $G^{ia}$ are $\ord{N_c}$, the GB-baryon coupling is $\ord{\sqrt{N_c}}$. This strong coupling at large $N_c$ demands the constraints of $SU(6)$, which will allow for $N_c$ consistency at higher orders in the effective theory. The third term gives the $SU(3)$ singlet mass splittings between baryons of different spins, and it is $\ord{p^0/N_c}$. The fourth term gives the contributions of quark masses to the baryon masses,   it is $\ord{p^2 N_c}$ and  gives $SU(3)$ breaking effects which are $\ord{p^2 N_c^0}$. This indicates a first issue with the interchange of chiral and large $N_c$ limits. As it becomes evident at the NLO due to the non-analytic terms of loop corrections, the limits do not commute, and for that reason it becomes necessary to make a choice: the choice made here is that $1/N_c$ is counted as a quantity of order $p$: $1/N_c=\ord{p}=\ord{\xi}$, which is coined as the $\xi$-expansion. The Lagrangian is now organized in powers of $\xi$. If   the $N_c$ dependencies of the matrix elements of the spin-flavor operators  are disregarded,   ${\cal{L}}_\B^{(1)}$ is $\ord{\xi}$.

   The construction of higher order Lagrangians  is  accomplished  making use of  the tools provided in Appendices~\ref{app:Algebra} and \ref{app:buildblocks}.  In this work   the Lagrangians of $\ord{\xi^2}$ and $\ord{\xi^3}$ are needed.
    Throughout, the spin-flavor operators appearing in the effective Lagrangians will be scaled by the appropriate powers of $1/N_c$ in such a way that all   LECs are of zeroth order in $N_c$. 
  The $1/N_c$ power  of a  Lagrangian term  with $n_\pi$ pion fields is given by \cite{Dashen:1994qi}: 
$n-1-\kappa+\frac{n_\pi}{2}$,
where the spin-flavor operator is $n$-body ($n$ is the number of factors of $SU(6)$ generators appearing in the operator),  and $\kappa$ takes into account  the $N_c$ dependency of the spin-flavor matrix elements.   The last term, ${n_\pi}/{2}$, stems from the factor $(1/F_\pi)^{n_\pi}$  carried by any term with $n_\pi$ GB fields.

For convenience the following definitions are used:
 \bea
  \delta \hat m&\equiv&  \frac{C_{\rm HF}}{N_c}{ \hat{S}^2}-\frac{c_1}{2\Lambda}\; \hat\chi_+\nonumber\\
  i\tilde D_0&\equiv& iD_0-\delta \hat m.
  \label{eq:deltam}
  \eea
  Note that $\delta \hat m$ gives rise to mass splittings between baryons which are $\ord{1/N_c}$ or $\ord{p^2}$.
  
With this,  the $\ord{\xi^2}$ Lagrangian is given by \footnote{The notation for the LECs used here differs from the ones used in ordinary BChPT due to the unifitcation of terms demanded by the $1/N_c$ expansion. The notation aims at distinguishing classes of terms in the Lagrangian, e.g.,   spin-independent mass  terms, spin-dependent mass terms, axial-vector couplings, etc. The identification of some of the LECs with those used in ordinary versions of BChPT are straigtforward.}:
 \bea
 { \cal{L}}_\B^{(2)}&=&\Bdag\left(    
( -\frac{1}{2 N_c m_0}+\frac{w_1}{\Lambda})\vec{D}^2 \right. +( \frac{1}{2 N_c m_0}-\frac{w_2}{\Lambda})\tilde{D}_0^2+
\frac{c_2}{\Lambda}\,  \chi^0_+
  \nonumber\\
   &+&  \frac{C_1^A}{N_c} u^{ia} S^i T^a+\frac{C_2^A}{N_c} \eps^{ijk}u^{ia}\{S^j,G^{ka}\}
   \nonumber\\
  &+&\kappa_0\; \eps^{ijk} F_{+ij}^0S^k+\kappa_1\;\eps^{ijk} F^a_{+ij}G^{ka}
   + \rho_0 F_{-0i}^0S^i+\rho_1 F^a_{-0i}G^{ia}
  \nonumber\\
  &+&   \frac{\tau_1}{N_c} u_0^a G^{ia} D_i + \frac{\tau_2}{N_c^2} u_0^a S^{i}T^{a} D_i + \frac{\tau_3}{N_c} \nabla_i u_0^a S^{i}T^{a}  + \tau_4 \nabla_i u_0^a G^{ia}  +  \cdots \bigg) \B,
  \label{eq:L2}
 \eea
 where  additional terms not explicitly displayed are not needed in the present work.
 Note that there are also $\ord{\xi^2}$ terms stemming from the $1/N_c$ suppressed terms in the LECs of the lower order Lagrangian. Similar comments apply to the higher order Lagrangians. Such terms require knowledge of the physics at $N_c>3$ to be determined, which can in principle be obtained  using LQCD results at varying $N_c$ \cite{DeGrand:2012hd,Cordon:2014sda}.
  
   Similarly,  the $\ord{\xi^3}$ Lagrangian needed here is given by:  
  \bea
 { \cal{L}}_\B^{(3)}&=&\Bdag \Big (\frac{c_3}{N_c\,\Lambda^3} \;\hat\chi_+^2+\frac{h_1 \Lambda}{N_c^3} \hat S^4+ \frac{h_2}{N_c^2\Lambda} \hat\chi_+\hat{S}^2+ \frac{h_3}{N_c\Lambda} \chi^0_+ \hat{S}^2+\frac{h_4}{N_c\, \Lambda} \; \chi_+^a\{S^i,G^{ia}\}
 \nonumber\\
  & +&\frac{C_3^A}{N_c^2} u^{ia} \{\hat S^2,G^{ia}\}+\frac{C_4^A}{N_c^2} u^{ia} S^i S^j G^{ja}\nonumber\\
  &+&\frac{D_1^A}{\Lambda^2} \chi_+^0 u^{ia} G^{ia}+\frac{D_2^A}{\Lambda^2} \chi_+^a u^{ia} S^{i}+\frac{D_3^A(d)}{\Lambda^2} d^{abc} \chi_+^a u^{ib} G^{ic} +\frac{D_3^A(f)}{\Lambda^2} f^{abc} \chi_+^a u^{ib} G^{ic} 
 \nonumber\\
  &+ & g_E \; [D_i,F_{+i0}]
+\alpha_1 \frac{i}{N_c} \eps^{ijk} F^a_{+0i} G^{ia} D_k
+\beta_1\frac{i}{N_c}  F^a_{-ij} G^{ia} D_j+\cdots\Big ) \B
\label{eq:L3}
 \eea

 In this work,  some terms $\ord{\xi^4}$ are needed for subtracting UV divergencies, but they are beyond the order of the present calculations and can be    consistently   eliminated. 
Through the calculation of the one-loop corrections to the self energies and the vector and axial vector currents, the $\beta$ functions associated with the LECs that affect  those  quantities are determined.

The terms in the effective Lagrangian  are constrained in their  $N_c$ dependence by the requirement of the consistency of QCD at large $N_c$. This     constraint  is in the form of a lower bound in the power in $1/N_c$ for each term  in the Lagrangian.     This leads in particular   to   constraints  on the $N_c$ dependencies of the ultra-violet (UV) divergencies of loop corrections,  which have to be subtracted by the corresponding counter-terms in the Lagrangian. 
The UV divergencies are necessarily polynomials in low momenta $p$ (derivatives), in $\chi_\pm$ and other sources, and in $1/N_c$ (modulo factors of $1/\sqrt{N_c}$ due to   $1/F_\pi$ factors in terms where GBs  are attached). Therefore, the structure of counter-terms is independent of any linking between the $1/N_c$ and chiral expansions. For this reason, in order to determine the UV divergencies,   the large $N_c$ and low energy limits can be taken independently. 
For a connected diagram with $n_B$ external baryon legs, $n_\pi$ external GB legs, $n_i$ vertices of type $i$ which have $n_{B_i}$ baryon legs and $n_{\pi_i}$  GB legs, and $L$ loops, the following topological relations hold \cite{Weinberg:1995mt,Weinberg:1991um}:
\bea
L=1+I_\pi+I_B-\sum n_i ,~~~~2 I_B+n_B=\sum n_i \;n_{B_i},~~~~2 I_\pi+n_\pi=\sum n_i \;n_{\pi_i},
\label{eq:topol}
\eea
where $I_\pi$ is the number of  GB propagators and $I_B$ the number of baryon propagators.

The chiral or low energy    order of a diagram, where  $\nu_{p_i}$ is the   chiral power of the vertex of type $i$, is   then given by \cite{Weinberg:1991um}:
\beq
\nu_p = 2 - \frac{n_B}{2} + 2 L +\sum_i n_i\;(\nu_{p_i} + \frac{n_{B_i}}{2} -2),
\label{eq:chiral-counting}
\eeq
Note that  $n_{B_i}$ is equal to 0 or 2 in the single baryon sector.
 
 On the other hand, the $1/N_c$ power of a connected diagram is determined by looking only at the vertices: the order in $1/N_c$ of a vertex of type $i$ is given  by: $\nu_{O_i}+\frac{n_{\pi_i}}{2}$, where $\nu_{O_i}$ is the order of the spin-flavor operator. Thus, the $1/N_c$ power of a diagram, upon use of the third Eq. \eqref{eq:topol}, is given by:
\beq
\nu_{\frac{1}{N_c}} =\frac{n_\pi}{2} +I_\pi+\sum{n_i\; \nu_{O_i}}~,
\label{eq:Nc-counting}
\eeq
where $n_\pi$ is the number of external pions, and $\nu_{O_i}$ the $1/N_c$ order of the spin-flavor operator of the  vertex of type $i$. Since $\nu_{O_i}$ can be negative (due to factors of $G^{ia}$ in vertices), there are individual diagrams with  $\nu_{\frac{1}{N_c}}$ negative and violating large $N_c$ consistency.  When the latter occurs, there must be other diagrams that cancel those violating terms. This will be clearly seen in the calculations presented here.
 
One can determine now the nominal counting of the one-loop contributions to the baryon masses and   currents. The LO baryon masses are $\ord{N_c}$, with hyperfine mass splittings that are $\ord{1/N_c}$ and $SU(3)$ symmetry breaking mass splittings that are $\ord{p^2}$. The one-loop correction shown in Fig.~\ref{fig:self-energy} has: $(L=1,~ n_B=2,~ n_\pi=0, ~n_1=2,~ \nu_{O_1}=-1,~n_{B_1}=2,~\nu_{p_1}=1)$ giving $\nu_p=3$ as it is well known, and  $\nu_{\frac{1}{N_c}}=-1$. Since there is only one possible diagram, this will be consistent if it contributes  $\ord{N_c}$ to the spin-flavor singlet component of the masses, it must contribute at $\ord{1/N_c}$ or higher to the hyperfine splittings, and at $\ord{N_c^0}$ to $SU(3)$ breaking. Indeed, this will be shown to be the case.  For the vector and axial-vector currents   the  one-loop diagrams are depicted in Figs.~\ref{fig:1-loop-VC} and \ref{fig:1-loop-AC} respectively.  Taking as example the axial currents,  at tree level it is $\ord{N_c}$, and the sum of the diagrams cannot scale as a higher power of $N_c$. Performing the counting for the individual diagrams one obtains:
$\nu_p(j)=2$ for $j=1,\cdots,4$, and $\nu_{\frac{1}{N_c}}(j)=-2$, $j=1,2,3$ and $\nu_{\frac{1}{N_c}}(4)=0$. Thus a cancellation must occur of the $\ord{N_c^2}$ terms when the contributions to the axial currents by the different diagrams are added, as it will be shown to be the case.
 
One can consider the case of two-loop diagrams, in particular  diagrams where the same GB-baryon  vertex Eq.(\ref{eq:Lagrangian-LO}) appears four times.  For the self energy the chiral power is $\nu_p(j)=5$, and individual diagrams give $\nu_{\frac{1}{N_c}}=-2$.  Thus a cancellation among the different diagrams must therefore occur. A comment is here in order: in Refs. \cite{CalleCordon:2012xz,Cordon:2014sda} the wave function renormalization factor was included in defining the baryon mass, but that is not   correct as in includes an incomplete inclusion of the two-loop contributions. In all cases, and as shown in this work, the diagrams that invoke the wave function renormalization factors play a key role in such cancellations.

Using the linked power counting $\xi$, $\ord{1/N_c}=\ord{p}=\ord{\xi}$,   the  $\xi$ order of a given Feynman diagram will then be   equal to  $\nu_p+\nu_{\frac{1}{N_c}}$ as given by Eqs.\eqref{eq:chiral-counting} and  \eqref{eq:Nc-counting}, which upon use of the topological formulas Eq.\eqref{eq:topol} leads to:
\beq
\nu_\xi=1+3 L+\frac{n_\pi}{2} +\sum_i n_i\;(\nu_{O_i}+\nu_{p_i}-1).
\label{eq:xi-counting}
\eeq
  The $\xi$-power counting of the UV divergencies is obvious from the earlier discussion. At one-loop  the masses have $\ord{\xi^2}$ and $\ord{\xi^3}$ counter-terms, while the axial currents will have $\ord{\xi}$ and $\ord{\xi^2}$ counter-terms. To two loops there are  in addition  $\ord{\xi^4}$ and $\ord{\xi^5}$, and  $\ord{\xi^3}$ and $\ord{\xi^4}$ counter-terms for masses and axial currents respectively. The non-commutativity of limits is manifested in the finite terms where the GB masses and/or momenta,  and $\delta \hat{m}$ appear combined in non-analytic terms, and are therefore sensitive to the linking of the two expansions. The $\xi$ expansion corresponds to not expanding such terms at all.

\section{Baryon masses}
\label{sec:Masses}

 In this section the baryon masses are analyzed to order $\xi^3$, or next-to-next to leading order  (NNLO), in the limit of exact isospin symmetry. To that order one must include the one-loop contribution  depicted in Fig. \ref{fig:self-energy} with the vertices from ${ \cal{L}}_\B^{(1)}$ given in Appendix \ref{app:buildblocks}. The contribution to the self-energy  is then given by:
\bea
\delta\Sigma_{1-loop}&=&i\,\frac{ \mathring{g}_A^2}{F_\pi^2}\;\sum_{a=1}^8\sum_{n} G^{ia} {\cal{P}}_n
G^{ia}\;\frac{\Gamma(1-\frac{d}{2})}{(4\pi)^{\frac d2}} \;J(1,0,M_a^2-(p_0-\delta m_n)^2,1,p_0-\delta m_n), ~~~
\label{eq:delta-Sigma}
%\\
\eea
where $n$ indicates the possible intermediate baryon  states in the loop,  ${\cal{P}}_n$ are the corresponding spin-flavor projection operators, the loop integral $J$ is given in Appendix \ref{app:loopintegrals}, $\delta m_n$ is the residual mass of the baryon in the propagator, i.e. $\delta \hat m$  in Eq. (\ref{eq:deltam}) evaluated for that state $n$,  $M_a$ is the mass of the Goldstone Boson in the loop (throughout the Gell-Mann-Okubo (GMO) mass relation $M_\eta^2= (4 M_K^2 - M_\pi^2)/3$ is used), and $p_0$ is the energy of the external baryon.  In the $\xi$ expansion,   the $SU(3)$ breaking effects   in $\delta m_n$   are  $\ord{\xi^2}$, and thus they can be neglected, i.e., one can simply use   $\delta \hat m\to \frac{C_{\rm HF}}{N_c} \hat S^2$ which is $\ord{\xi}$. 
In the specific evaluation of $\delta\Sigma_{1-loop}$ for a given  baryon state denoted by $"in"$, $p_0=\delta m_{in}+\p$, where $\p$ is the kinetic energy  $\ord{p^2/N_c}$.  The non-commutativity of the $1/N_c$ and  chiral expansions of course resides in the non-analytic terms of the loop integral  through their dependence  on the ratios  of the small scales $(\delta m_n-\delta m_{in})/M_a$. Notice that when the one-loop integrals are written in terms of the residual momentum $\p$, they do not depend on the spin-flavor singlet piece of $\delta \hat m$. $\p$ is naturally associated with $i\tilde{D}^0$.
The one-loop contribution to the wave function renormalization factor is given by:
$\delta Z_{1-loop}=\left.\frac{\partial}{\partial \p}\delta\Sigma_{1-loop}\right\arrowvert_{{\p\to 0}}$.
Appendices~\ref{app:Algebra} and ~\ref{app:opred} provide all the necessary  elements  for the evaluation of the spin-flavor matrix elements in Eq.~\eqref{eq:delta-Sigma}.
The explicit final expressions for the self energy are straightforwardly calculated using those elements, and are not given explicitly because they are too lengthy.

The correction to the baryon mass is given by setting $\p=0$ in the self-energy correction, and the  mass of the baryon  state $\mid \! S, Y I\rangle$ then reads:
\beq
m_\B(S,Y,I)=N_c m_0+\frac{C_{\rm HF}}{N_c} S(S+1)-\frac{c_1}{2\Lambda}((N_c+2{\cal{S}}) M_\pi^2-2 {\cal{S}} M_K^2)+\delta m^{1-loop+CT}_\B(S,Y,I),
\label{eq:baryon-mass}
\eeq
where ${\cal{S}}$ is the strangeness,  $\delta m^{1-loop+CT}_\B(S,Y,I)$  is the contribution from the one-loop diagram in Fig.~\ref{fig:self-energy} and CT  denotes  counter-term contributions. From both types of contributions, there are $\ord{\xi^2}$ and $\ord{\xi^3}$ terms,  and the calculation is exact to the  latter order, as can be deduced from the previous discussion on power counting. 
 Note that in LO the LEC $C_{\rm HF}$ is equal to the   hyperfine splitting $M_\Delta-M_N$ in the real world $N_c=3$.

\begin{center}
\begin{figure}[h]
\centerline{\includegraphics[width=8.cm,angle=-0]{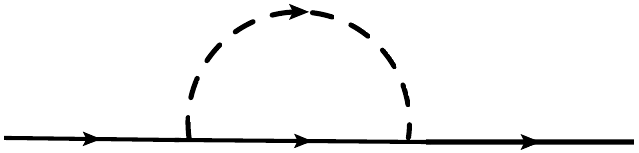}}
\caption{One-loop contribution to baryon self energy. }
\label{fig:self-energy}
\end{figure}
\end{center}
\vspace{-1cm}

The   ultraviolet divergent pieces of the self energy can be brought to have the following form:   
\bea
\delta\Sigma_{1-loop}^{UV}
&=&\frac{\lambda_\eps}{(4 \pi)^2}
\left(\frac{\mathring{g}_A
}{F_\pi}\right)^2
\left( \p M_a^2 G^{ia}G^{ia}+\frac{1}{2} M_a^2[[\delta \hat m,G^{ia}],G^{ia}]-\left.\frac{2}{3}{\p}^3\right. \right.\\
&-&\left.{\p}^2[[\delta \hat m,G^{ia}],G^{ia}]-\p[[\delta \hat m,[\delta \hat m,G^{ia}]],G^{ia}]-\frac{1}{3}[[\delta \hat m,[\delta \hat m,[\delta \hat m,G^{ia}]]],G^{ia}]\right),\nonumber
\label{eq:UVmass}
\eea
 where $\lambda_\eps\equiv 1/\eps-\gamma+\log 4\pi$.   Using the $SU(3)$ singlet and octet components of the quark masses, $m^0$ and $m^a$,   the meson mass-squared matrix can be written as:
 \beq
 {M^2}^{ab}=2B_0(\delta^{ab} m^0+\frac{1}{2}d^{abc}m^c),
 \eeq
 and therefore,
 \beq
 M_a^2 W^{aa}={M^2}^{ab} W^{ab},
 \eeq
for any symmetric $\bf{8\times 8}$ tensor $W$. In terms of $M_\pi$ and $M_K$ one	 has: $m^0 =\frac{1}{3}(2\hat m+m_s)= \frac{2 M_K^2 + M_\pi^2}{6 B_0}$ and  
 $m^a =\delta^{8a}\frac{2}{\sqrt{3}}(\hat m-m_s)=\delta^{8a}\frac{  2 (-M_K^2 + M_\pi^2)}{\sqrt{3} B_0}$.

In order to obtain from Eq.(\ref{eq:UVmass}) the counter-terms necessary to renormalize the mass and wave function, one uses the results in Appendix \ref{app:opred}.
The explicit UV divergent  and   polynomial  (in $1/N_c$, $m_q$,  $\p$)  terms of the self energy   are the given by:
\bea
\delta\Sigma^{\text{poly}}&=&-
\frac{1}{(4\pi)^2} 
\left( \frac{\mathring g_A}{F_\pi} \right)^2\left\{  \left(\frac 73+\lambda_\eps\right) B_0 \frac{C_{\rm HF}}{N_c} \left(\left(\frac 34N_c(N_c+6)-7 \hat S^2\right) m^0\right.\right.\nonumber\\
&+&\left.\left(-2\{S^i,G^{ia}\}+\frac 34(N_c+3) T^a\right)m^a\right)       \nonumber\\
&+&(\frac 83+\lambda_\eps)\frac{C_{\rm HF}^3}{N_c^3}\left(-N_c(N_c+6)+\frac 13(36-5N_c(N_c+6))\hat S^2   +12 \hat S^4\right)         \nonumber\\
&+&\p\left( (1+\lambda_\eps) B_0\left(\left(-\frac 38N_c(N_c+6)+\frac 56 \hat S^2\right)m^0\right.+ \left(\frac{7}{12}\{S^i,G^{ia}\}-\frac 38(N_c+3) T^a\right)m^a\right)\nonumber\\
&+&\left. \left.(2+\lambda_\eps)\frac{C_{\rm HF}^2}{N_c^2}\left(\frac 32 N_c(N_c+6)+(-18+N_c(N_c+6))\hat S^2-4 \hat S^4\right)\right)\right\},
 \eea
 where terms of higher powers in $\p$ have been disregarded.   A few observations on $\delta\Sigma^{\text{poly}}$ are in order: 1) the  contributions to the spin-flavor singlet component of the masses  is $\ord{p^2 N_c^0}$ and proportional to $C_{\rm HF}$,  the spin-symmetry breaking is  $\ord{1/N_c^2}$, and the $SU(3)$ breaking is $\ord{p^2/N_c}$; 2) the  UV divergencies in the mass  are produced by the contribution of the partner baryon in the loop, i.e. baryon of different spin, and is therefore determined by the mass splitting, i.e., by $C_{\rm HF}$;
3)  the contributions to $\delta Z$ are suppressed by powers of $1/N_c$, but with two exceptions, namely, there is a spin-flavor singlet contribution proportional to $m^0$ which is $\ord{N_c}$ and a term proportional to $m^a$ which is $\ord{N_c^0}$. The term $\ord{N_c}$ in $\delta Z$  is of key importance for the mechanism of cancellations of $1/N_c$ power counting violating terms, as it is shown later in the analysis of the one-loop contributions to the currents.

The counter-terms  for renormalizing  the masses and wave functions are $\ord{\xi^2}$ and $\ord{\xi^3}$ (all contributions $\ord{\xi^4}$ are  consistently dropped)   and involve terms that appear in $  { \cal{L}}_\B^{(1)}$ with higher order terms in $1/N_c$ in the LECs  and    terms in   ${ \cal{L}}_\B^{(2,3)}$.   To renormalize,  the LECs are written as: $X=X(\mu)+\frac{1}{(4\pi)^2}\beta_X \lambda_\eps$, where $\mu$ is the  renormalization scale and
the beta-functions  $\beta_X$ necessary to renormalize the masses are given in Table \ref{tab:betafunctionsmass}. The reader can easily work out the renormalization of the wave functions.

\begin{centering}
\begin{table*}[ttt]
{
\begin{tabular}{|c|c| }
\hline\hline
LEC & $ F_\pi^2  \beta/  g_A^2 $     \\\hline
$m_0$ &$- \frac{N_c+6}{N_c^3}C_{\rm HF}^3$   \\
$C_{{\rm HF}}$ & $\frac{36-5N_c(N_c+6)}{3N_c^2}C_{\rm HF}^3 $  \\
$c_1$ & $-\frac 38\frac{N_c+3}{N_c} \Lambda C_{\rm HF}$  \\
$c_{2}$ & $ \frac{3}{16}(2N_c+9)  \Lambda C_{\rm HF} $       \\
$c_{3}$ & 0       \\
$h_1$ & $-\frac{12}{\Lambda} C_{\rm HF}^3$   \\
$h_2$ & $0$   \\
$h_3$ & $\frac 74 \Lambda C_{\rm HF}$   \\
$h_4$ & $\frac 12 \Lambda C_{\rm HF}$   \\
\hline\hline 
\end{tabular}
}
\caption{$\beta$ functions for mass renormalization 
}  
\label{tab:betafunctionsmass}
\end{table*}
\end{centering}
Finally the non-analytic contributions to $\delta \Sigma$ are:
\bea
\delta\Sigma^{\text{NA}}&=&-\frac{1}{(4\pi)^2} \left( \frac{\mathring g_A}{F_\pi} \right)^2 \sum_{n} G^{ia}{\cal{P}}_n G^{ia}\nonumber\\
&\times& \Bigg((p_0-\delta m_n)(M_a^2-\frac 23(p_0-\delta m_n)^2)\log\frac{M_a^2}{\mu^2} \\
&+&\left. \frac 23(M_a^2-(p_0-\delta m_n)^2)^{\frac 32}(\pi+2\arctan\left( \frac{p_0-\delta m_n}{\sqrt{M_a^2-(p_0-\delta m_n)^2} }\right)\right).\nonumber
\eea

At tree level, and up to order $\xi^3$, baryon masses satisfy the GMO and Equal Spacing (ES) relations, which   hold unchanged at arbitrary $N_c$. The deviations from these relations are given by the non-analytic terms in the self energy, i.e., they are  calculable to the one-loop order,  and in the strict large $N_c$  limit they are  $\ord{p^3/N_c}$ and  $\ord{p^2/N_c^2}$. The calculated deviations  compare to   the observed ones   as follows: GMO:  $(3 m_\Lambda+m_\Sigma)-2(m_N+m_\Xi)= \Delta_{GMO}=$  Th: $(g_A^N/F_\pi)^2\times 2.42 \;10^5\;\text{MeV}^3$ vs  Exp:  $25.8$ MeV, and ES:  $m_{\Xi^*}-2m_{\Sigma^*}+m_\Delta= \Delta_{ES}=$ $(g_A^N/F_\pi)^2\times (-3.72 \;10^4)\; \text{MeV}^3$ vs $  -4\pm 7$ MeV, where for the theoretical evaluation   $C_{\rm HF}=m_\Delta-m_N$ was used.   Note that using the physical $g_A^N=1.267\pm0.004$ and $F_\pi=93$ MeV, the value of  $\Delta_{GMO}$ turns out to be significantly larger than the physical one. When studying the axial couplings, it will be found that the     LO        value of the axial coupling   is smaller than the physical one. In fact,  $\Delta_{GMO}$ could be used in determining  the  ratio $g_A^N/F_\pi$ at LO.  Expanding   $\Delta_{GMO}$   in the strict large $N_c$ limit one obtains:
\bea
\Delta_{GMO}&=&- \left( \frac{\mathring{g}_A}{4\pi F_\pi} \right)^2\left(\frac{2\pi}{3}\left(M_K^3-\frac 14 M_\pi^3-\frac{2}{\sqrt{3}} (M_K^2-\frac 14 M_\pi^2)^{\frac 32}\right)\right.\nonumber\\
&+& \left.\frac{C_{\rm HF}}{2N_c}\Big(4M_K^2\log\left( \frac {4M_K^2- M_\pi^2}{3M_K^2}\right)-  M_\pi^2 \log\left(  \frac {4M_K^2-\frac 13 M_\pi^2}{3M_\pi^2}\right)\Big)\right)\nonumber\\
&+&\ord{1/N_c^3}.
\eea
For the physical $M_K$ and $M_\pi$    the shown expansion is within 30\% of  the exact result, and the expansion gives a good approximation  for $N_c>5$.
 Note the large cancellations that appear within the first line and within the second line of the equation, and also the tendency to cancel between the first and second lines.   In the physical case and not expanding in $1/N_c$  it is found that the numerical dependency of  $\Delta_{GMO}$ on $C_{\rm HF}$ is not very significant. One also observes that  only 43\%  of $\Delta_{GMO}$ is contributed by the octet baryons in the loop, and thus the decuplet contribution is very important.  $\Delta_{GMO}$ is therefore an important observable for assessing whether the decuplet baryons ought to be  included  or not in the effective theory; as indicated earlier, this however depends on the value the LO $\mathring g_A$, which to be independently determined  requires the analysis of other observables, namely  the axial currents.  Along the same lines     $\Delta_{ES}$ can be analyzed, although in this case the experimental uncertainty is rather large.

 Disregarding  the term proportional to $h_2$ in $ { \cal{L}}_\B^{(3)}$ Eq.(\ref{eq:L3}), which gives $SU(3)$ breaking in the hyperfine splittings,   one additional relation follows, first found by G\"ursey and Radicati \cite{Gursey:1992dc}, namely:
\beq
\Delta_{GR}=m_{\Xi^*}-m_{\Sigma^*}-(m_{\Xi}-m_{\Sigma})=0,~~~~\text{Exp:  } 21\pm 7 ~\text{MeV}\label{GRRel},
\eeq
which relates $SU(3)$ breaking in the octet and decuplet, and which is valid  for arbitrary $N_c$.
The deviation from that relation (\ref{GRRel}) is due to $SU(3)$ breaking effects in the hyperfine interaction that splits $\bf 8$ and $\bf 10$ baryons, and  such deviation starts with the term proportional to $h_2$ which is  $\ord{p^2/N_c}$. In addition the one-loop contributions to it are free of UV divergencies and the non-analytic terms when expanded in the large $N_c$ limit give contributions  $\ord{1/ N_c^2}$. To one-loop:
\bea
\Delta_{GR}&=&
 \frac{h_2}{\Lambda} \frac{12}{N_c}M_K^2
 + \left( \frac{\mathring g_A}{4\pi F_\pi} \right)^2 \left(  \frac{2\pi}{9} M_K^3+\frac{(9N_c-43)\pi}{72}\left(M_K^2-\left(\frac{3 C_{\rm HF}}{N_c}\right)^2\right)^{\frac 32}   \right.\nonumber\\
&-&  \frac{N_c-3}{24}    \Big[  3 \left(M_K^2-\left(\frac{5 C_{\rm HF}}{N_c}\right)^2\right)^{\frac 32} \Big(\pi-2\arctan   \frac{5 C_{\rm HF}}{N_c \sqrt{M_K^2-\left(\frac{5 C_{\rm HF}}{N_c}\right)^2}} \Big)   \nonumber\\
&+& \left.\left.10  \left(M_K^2-\left(\frac{3 C_{\rm HF}}{N_c}\right)^2\right)^{\frac 32} \arctan  \frac{3 C_{\rm HF}}{N_c \sqrt{M_K^2-\left(\frac{3 C_{\rm HF}}{N_c}\right)^2}} +\frac{240}{N_c^3} C_{\rm HF}^3 \log M_K^2\Big]\right)\right.\nonumber\\
&-& \left(M_K \rightarrow M_\pi \right)\nonumber\\
&=&  \frac{h_2}{\Lambda} \frac{12}{N_c}(M_K^2-M_\pi^2)+\frac{3\pi}{N_c} \left( \frac{\mathring g_A C_{\rm HF}}{4\pi F_\pi} \right)^2 (M_K-M_\pi)+\ord{\frac{\log (M_K/M_\pi)}{N_c^3}},
\eea
where the last line corresponds to strictly expanding in the large $N_c$ limit. For the physical $M_\pi$, $M_K$, and $C_{\rm HF}$, the $1/N_c$ expansion  of  $\Delta_{GR}$ is  however   only reasonable for   $N_c>8$: clearly the non-analytic dependency in $1/N_c$ is important,   showing the need for the combined $\xi$ expansion in the physical case, similarly to what occurs for $\Delta_{GMO}$.
Still, the  understanding of the smallness of the deviation is connected with the $1/N_c$ expansion. Finally, it is important to emphasize, as indicated earlier,  that all the relations are   not explicitly dependent  on $N_c$, and their deviations are suppressed by powers of $1/N_c$ at large $N_c$.

The $\sigma$-terms  are obtained following the Hellman-Feynman theorem, $\sigma_{B m_q}\equiv m_q \partial m_B/\partial m_q$, where   $m_q$ can be taken to be $\hat m,\;m_s$, or the $SU(3)$ singlet and octet components of the quark masses, namely   $m^0=(2\hat m+m_s)/3$ and $m^8=2/\sqrt{3} (\hat m - m_s)$. Naturally they will satisfy the same relations discussed above for the masses. In particular $\sigma$-terms associated with the same $m_q$  are  related via those relations and their deviations are calculable as described before for the masses.  In addition to the GMO and ES relations, the following tree level  $\ord{\xi^3}$  relations hold,
\bea
 \;\sigma_{N m_s}&=& \frac{m_s}{8
   \hat m} (-4 (N_c-1)\;\sigma_{ 
   N \,\hat  m}+(N_c+3) \;\sigma_{ \Lambda \,\hat  m}+3 (N_c-1) \;\sigma_{ \Sigma \,\hat  m})
   \nonumber\\ \;\sigma_{ \Lambda \,m_s}&=& \frac{m_s}{8
   \hat m} (-4 (N_c-3) \;\sigma_{
   N\,\hat  m}+(N_c-5) \;\sigma_{ \Lambda \,\hat  m}+3 (N_c-1) \;\sigma_{ \Sigma \,\hat  m})
  \\ \;\sigma_{ \Sigma \,m_s}&=& \frac{m_s}{8
   \hat m}(-4 (N_c-3) \;\sigma_{
   N\,\hat  m}+(N_c+3) \;\sigma_{ \Lambda \,\hat  m}+(3 N_c-11) \;\sigma_{ \Sigma \,\hat  m})
   \nonumber\\ \;\sigma_{ \Delta \,m_s}&=& \frac{m_s}{8
   \hat m}(-4 (N_c-1) \;\sigma_{ \Delta
   \,\hat  m}-5 (N_c-3) (\;\sigma_{ \Lambda \,\hat  m}-\;\sigma_{ \Sigma \,\hat  m})+4 N_c \;\sigma_{
   {\Sigma ^*}\,\hat  m}) \nonumber\\ \;\sigma_{ {\Sigma ^*}\,m_s}&=&
  \frac{m_s}{8
   \hat m} (-(N_c-3) (4 \;\sigma_{ \Delta \,\hat  m}+5 \;\sigma_{ \Lambda \,\hat  m}-5 \;\sigma_{
   \Sigma\,\hat  m})+4  (N_c-2) \;\sigma_{ {\Sigma ^*}\,\hat  m}).\nonumber
\eea
Several of these relations   are poorly satisfied. The deviations are calculable and given by the non-analytic contributions to one-loop.  It is easy to understand why these relations receive large corrections:   they behave at large $N_c$ as $\ord{p^3 N_c}$. This implies that   tree level relations   used to relate $m_s$ and $\hat m$ $\sigma$ terms will in general receive    large non-analytic deviations.  In the physical case $N_c=3$, those deviations are numerically large for the first, third, and fourth relations above.  This in particular affects the nucleon strangeness $\sigma$ term, and thus indicates that its estimation from arguments based on tree level relations is subject to important corrections \cite{AlarconFernandoGoity}. 
 In terms of the octet components of the quark masses, in addition to GMO and ES relations one finds:
\bea
 \;\sigma_{ N \,m^8}&=& \frac{(N_c+3) \;\sigma_{ \Lambda \,m^8}+3
   (N_c-1) \;\sigma_{ \Sigma \,m^8}}{4 (N_c-3)}\\
   \;\sigma_{ \Delta
   \,m^8}&=& \frac{-5 (N_c-3) \;\sigma_{ \Lambda \,m^8}+5 (N_c-3)
   \;\sigma_{ \Sigma \,m^8}+4 N_c \;\sigma_{ \Sigma ^* \,m^8}}{4
   (N_c-3)} ,
  \eea
  where it can be readily  checked  that they are well defined for $N_c\to 3$ as  the numerators on the  RHS are proportional to $(N_c-3)$.  These relations are violated at large $N_c$ as  $\ord{p^3 N_c^0}$. For both relations in the limit $N_c\to\infty$ one finds $\text{LHS}-\text{RHS}=\frac{N_c}{128\pi}\left(\frac{\mathring g_A}{F_\pi}\right)^2 (M_K-M_\pi)(M_K^2-M_\pi^2)+\ord{1/N_c}$. Thus they are not as precise as the GMO and ES relations.

Finally,  if the LEC constant $h_3$ vanishes,   one extra tree-level relation related to Eqn. (\ref{GRRel}) follows, namely, 
\bea
\sigma_{\Xi^* m^8}-\sigma_{\Sigma^* m^8}-(\sigma_{\Xi m^8}-\sigma_{\Sigma m^8})&=&0
\eea
which is   only violated at large $N_c$ as $\ord{1/N_c^2}$, and thus expected to be very good.

To complete this section, fits to the octet and decuplet baryon masses including results from LQCD are presented. This in particular allows for  exploring the range of validity of the calculation as the quark masses are increased.
The mass formula for the fit is  \footnote{ A useful formula for the  term  proportional to $h_4$ is \cite{Matagne:2006xx}: \\
$S^i G^{i8}= \frac{1}{\sqrt{3}} \left( \frac 34 \hat I^2-\frac 14 \hat S^2-\frac{1}{48} N_c(N_c+6)+\frac 18 (N_c+3)Y-\frac{3}{16} Y^2\right)=\frac{1}{16\sqrt{3}}(12 \hat I^2-4\hat S^2+3{\cal{S}}(2-{\cal{S}}))$, where ${\cal{S}}$ is the strangeness. This term is responsible for the tree-level mass splitting between $\Lambda$ and $\Sigma$.}
:
\bea
m_B&=& 
N_c m_0+\frac{C_{\rm HF}}{N_c}\hat S^2-\frac{c_1}{2 \Lambda} \hat\chi_+ -\frac{c_2}{\Lambda}\chi_+^0 -\frac{c_3}{N_c \Lambda^3}\hat \chi_+^2 \nonumber\\
&-&\frac{h_2}{N_c^2 \Lambda}\hat \chi_+ \hat S^2- \frac{h_3}{N_c \Lambda}\chi_+^0\hat S^2-2\frac{h_4}{N_c\Lambda}  \tilde \chi_+^a   S^i G^{ia}+\delta m_B^{1-loop} ,
\eea
where, in the isospin symmetry limit, $\chi_+^0\to 4B_0 m^0,~\tilde\chi_+^a\to 8B_0\delta^{a8}  m^8,~\text{and }\hat \chi_+\to 4B_0(m^8 T^8+N_c m^0)$.
The fits at $N_c=3$ cannot  obviously give the $N_c$ dependence of LECs. LECs of terms that depend on quark masses can be more completely determined by fits that include the LQCD results for different quark masses, e.g., $c_2$ and the various $h's$. For this reason such combined fits are presented here, in Table \ref{tab:massfits} and in Fig. \ref{fig:BChPT-combined-fits} in Appendix \ref{app:LQCDmasses}. Also, some LECs are redundant at $N_c=3$, and are thus set to vanish for the fit. The constant $c_3$ is also set to vanish as it turns out to be of marginal importance for the fit. A test of mass relations is shown in Table \ref{tab:massrels}.
\begin{table}[h!]\small
\begin{tabular} {llllllll}
\hline\hline $\chi^2_{\rm dof}$ & $m_0\,\text{[MeV]}$ & $C_{\rm HF}\,\text{[MeV]}$ & $ ~~~~~ c_1$ &  $~~~~~c_{2}$& $~~h_2$ & $~~~h_{3}$  & $~~~h_4$   \\
\hline 0.47 & ~~221(26) & ~~215(46) &$ -1.49(1)$ & $-0.83(5)$& 0.03(3)& $0.61(8)$  & $0.59(1) $ \\
        0.64& ~~191(5) & $~~242(20)$ &$ -1.47(1)$ & $-0.99(3)$& 0.01(1) & $0.73(3)$ &   $0.56(1) $ \\  \hline \hline
\end{tabular}
\caption{Results for LECs: the ratio $\mathring g_A/F_\pi=0.0122\text{ MeV}^{-1}$ is fixed by using $\Delta_{GMO}$.  The first row is the fit to LQCD octet and decuplet baryon masses   \cite{Alexandrou:2014sha}  including results for $M_\pi\leq 303$ MeV (dof=50), and second row is the fit including also the physical masses (dof=58). Throughout the   $\mu=\Lambda=m_\rho$.} 
\label{tab:massfits}
\end{table}

\begin{table}[h!]\small
    \begin{tabular}{cc|cc|cc|cc|cc}
        \hline      \hline  $M_\pi$ & $M_K$ & \multicolumn{2}{c|}{$\Delta_{GMO}$ }&  \multicolumn{2}{c|}{$\Delta_{GR}$ } &  \multicolumn{2}{c|}{$\Delta_{ES1}$ }&  \multicolumn{2}{c}{$\Delta_{ES2}$ }\\
        \multicolumn{2}{c|} {[MeV]} & Exp/LQCD &~~ Th ~~& Exp/LQCD &~~ Th~~& Exp/LQCD & ~~Th~~& Exp/LQCD & ~~Th~~ \\
        \hline  139 & 497 & 31$\pm$42 & 46 & 23$\pm$30 & 38 & -6$\pm$30 & -14 & -9$\pm$30 & -14 \\
        213 & 489 & 75$\pm$70 & 33 & 0$\pm$72 & 29 & -40$\pm$97 & -11 & 9.2$\pm$83 & -11 \\
        246 & 499 & 124$\pm$77 & 30 & -7$\pm$75 & 25 & -46$\pm$101 & -11 & 23$\pm$86 & -11 \\
        255 & 528 & 133$\pm$89 & 37 & -12$\pm$94 & 26 & -32$\pm$125 & -14 & 29$\pm$108 & -14 \\
        261 & 524 & 139$\pm$99 & 35 & 24$\pm$103 & 25 & -29$\pm$138 & -13 & -3$\pm$119 & -13 \\
        302 & 541 & 77$\pm$87 & 32 & -14$\pm$94 & 23 & -30$\pm$125 & -13 & 46$\pm$108 & -13 \\
        \hline\hline
    \end{tabular}
    \caption{Deviations from mass relations in MeV. Here $\Delta_{\text{ES1}}=m_{\Xi^*}-2m_{\Sigma^*}+m_\Delta$ and  $\Delta_{\text{ES2}}=m_{\Omega^-}-2m_{\Xi^*}+m_{\Sigma^*}$. } 
    \label{tab:massrels}
\end{table}   

The study of the fits show that at fixed $M_K\sim 500$ MeV,  the physical plus LQCD results up to $M_\pi\sim 300$ MeV can be fitted  with natural size LECs. The LEC $h_2$ which enters in $\Delta_{GR}$ is best determined by fixing it using $\Delta_{GR}$ in the physical case, and then the rest of the LECs are determined by the overall fit.  In this way, the deviations of the mass relations are one of the predictions of the effective theory, and can therefore be used as a test of LQCD calculations.  At present the errors in the LQCD calculations are relatively large, and thus such a test is  not  yet very significant.

 \section{Vector currents:  charges}
\label{sec:Vector}
In this section the one-loop corrections to the vector current charges are calculated.  The analysis is similar to that   carried out in \cite{Flores-Mendieta:2014vaa}, except that in that reference higher order terms in $1/N_c$ in the GB-baryon vertices were included.   In the $\xi$ expansion and the order considered here such higher order terms are not required.   At lowest order the charges  are simply given by  the generators $T^a$,  the one-loop corrections  are UV finite, and since up to $\ord{\xi^3}$ the  Ademollo-Gatto theorem (AGT) is satisfied,  the corrections to the charges are unambiguously given at one-loop. 

\begin{center}
\begin{figure}[h]
\centerline{
\includegraphics[width=12cm,angle=0]{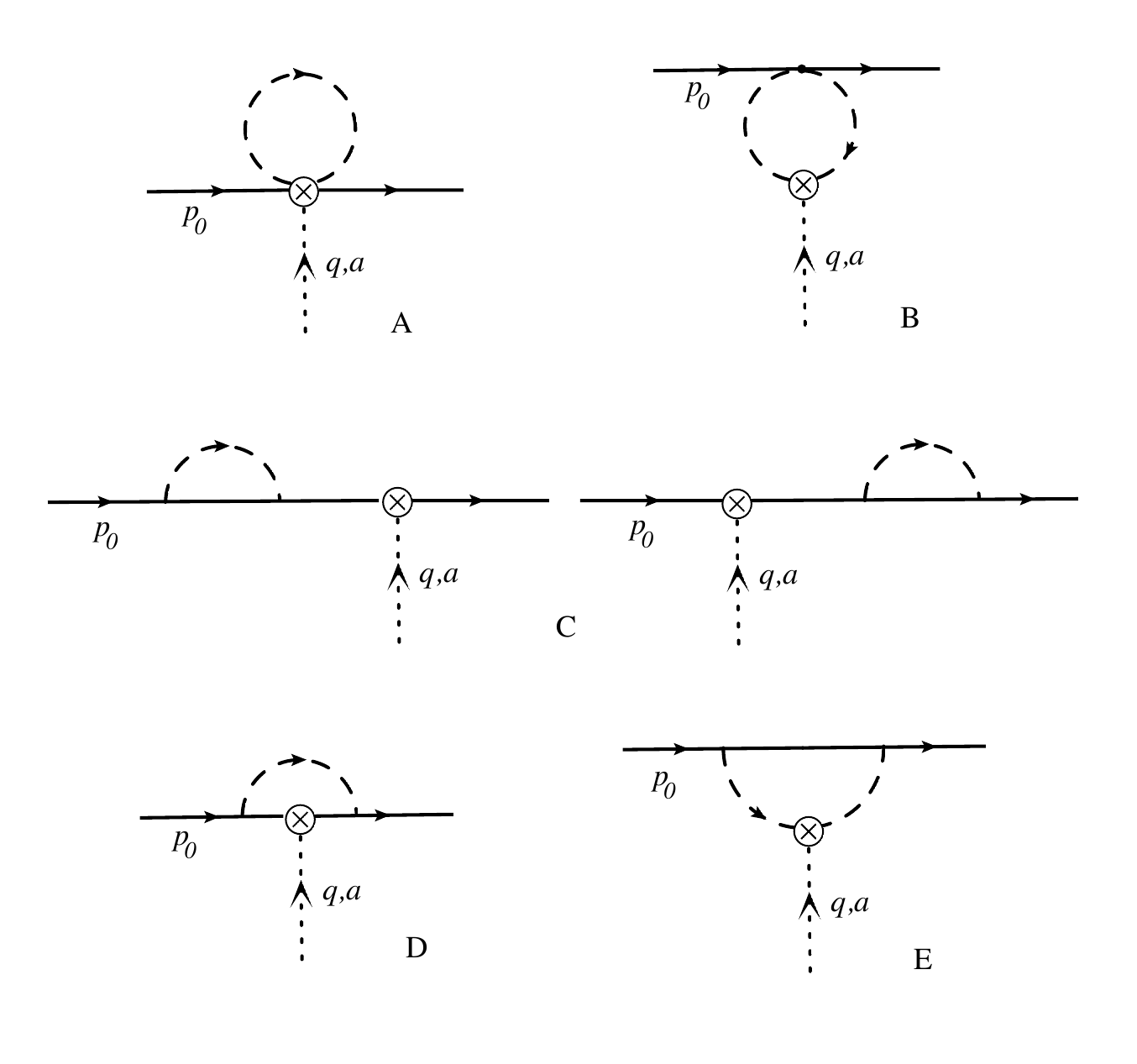}  }
\caption{Diagrams contributing to the 1-loop corrections to the vector charges.     }
\label{fig:1-loop-VC}
\end{figure}
\end{center}

The one-loop diagrams are shown in Fig. \ref{fig:1-loop-VC}, and the corrections to the charges are obtained by evaluating the diagrams at $q\to 0$. In that limit the UV divergencies as well as the finite polynomial terms in quark masses and $\delta \hat m$ cancel in each of the two sets of diagrams, $A+B$, and $C+D+E$, as required by the AGT. 
The results for the diagrams are the following:
\bea
 A &=& -\frac{i}{2F_\pi^2} f^{abc}f^{bcd} T^d I(0,1,M_b^2)\nonumber\\
 B &=&\frac{i}{4F_\pi^2} f^{abc}f^{bcd} T^d({q^0}^2 K(q,M_b,M_c)+4 q^0 K^0(q,M_b,M_c)+4 K^{00}(q,M_b,M_c)) \nonumber\\
 C &=&\frac 12 \{T^a,\delta \hat Z_{1-loop}\}\nonumber\\
 D &=&i \left( \frac{\mathring g_A}{F_\pi} \right)^2 \sum_{n_1,n_2} G^{ib} {\cal{P}}_{n_2} T^a  {\cal{P}}_{n_1} G^{jb}\;\frac{1}{q_0-\delta m_{n_2}+\delta m_{n_1}}\nonumber\\
&\times&\left(H_{ij}(p_0-\delta m_{n_1},M_b)-H_{ij}(p_0+q_0-\delta m_{n_2},M_b)\right)\nonumber\\
 E &=&\left( \frac{\mathring g_A}{F_\pi} \right)^2 f^{abc} \sum_{n} G^{ib}   {\cal{P}}_{n} G^{jc} H_{ij0}(p_0-\delta m_n, q, M_b,M_c),
\eea
where the integrals $K$, $K^\mu$, $K^{\mu\nu}$, $H_{ij}$ and $H_{ij0}$ are given in Appendix \ref{app:loopintegrals}. Since the temporal component of the current can only connect baryons with the same spin,   $q_0$ is equal to the $SU(3)$ breaking mass difference between them plus the kinetic energy transferred by the current, which are all $\ord{p^2}$, and  can be neglected:   the limit $q_0\to 0$ must then be taken in the end. Diagram $D$  indeed  requires  a careful handling of that limit  in the cases when the denominator vanishes. The same is the case for diagram  $F$ in the axial-vector currents in next section. The $U(1)$ baryon number current is used to check the calculation: only diagrams $C+D$ contribute, and as required cancel each other.

The UV divergent and polynomial pieces contributed by the diagrams are the following:
\bea
A^{\text{poly}}&=&\frac{\lambda_\eps+1}{(4\pi)^2} \frac{1}{2F_\pi^2}f^{abc}f^{bcd} M_b^2 T^d\nonumber\\
B^{\text{poly}}&=&-\frac{\lambda_\eps+1}{(4\pi)^2}  \frac{1}{2F_\pi^2} f^{abc}f^{bcd}  T^d(M_b^2+\frac 16\, {\vec{q}}^{\;2})\nonumber\\
C^{\text{poly}}&=&\frac{1}{(4\pi)^2}\left( \frac{\mathring g_A}{F_\pi} \right)^2 \frac{1}{2}\left\{T^a,(\lambda_\eps+1) M_b^2 G^{ib}G^{ib}-2(\lambda_\eps+2) G^{ib}[\delta \hat m,[\delta\hat m,G^{ib}]]\right\}\nonumber\\
D^{\text{poly}}&=&\frac{1}{(4\pi)^2}\left( \frac{\mathring g_A}{F_\pi} \right)^2 \frac{1}{3}\sum_{n_1,n_2}G^{ib}{\cal{P}}_{n_2}  T^a {\cal{P}}_{n_1} G^{ib} \frac{1}{q_0-\delta m_{n_2}+\delta m_{n_1}}
\nonumber
\\
&\times& \left\{(p_0-\delta m_{n_1})(3(\lambda_\eps+1)M_b^2-2(\lambda_\eps+2) (p_0-\delta m_{n_1})^2-\{p_0\to p_0+q_0,\delta m_{n_1}\to \delta m_{n_2}\})\right\}\nonumber\\
&=& \frac{1}{(4\pi)^2}\left( \frac{\mathring g_A}{F_\pi} \right)^2 \frac{1}{3}\left\{-3(\lambda_\eps+1) M_b^2 G^{ib}T^aG^{ib}\right.\nonumber\\
&+&\left. 2(\lambda_\eps+2)\left([\delta \hat m,[\delta\hat m,G^{ib}]]T^a G^{ib}+G^{ib} T^a [\delta \hat m,[\delta\hat m,G^{ib}]]-[\delta \hat m,G^{ib}] T^a [\delta \hat m,G^{ib}] \right)\right\}\nonumber\\
E^{\text{poly}}&=&-\frac{1}{(4\pi)^2}\left( \frac{\mathring g_A}{F_\pi} \right)^2 \frac{i}{6}f^{abc}\sum_{n} G^{ib}  {\cal{P}}_{n} G^{jc}
\left\{\lambda_\eps (2q^i q^j+q^2 g^{ij})+q^2 g^{ij}\right.\nonumber\\
&-& 3\left. g^{ij}((\lambda_\eps+1)(M_b^2+M_c^2)-(\lambda_\eps+2)(\delta m_{in}-2 \delta m_n+\delta m_{out})^2)\right\}\nonumber\\
&=& -\frac{1}{(4\pi)^2}\left( \frac{\mathring g_A}{F_\pi} \right)^2 \frac{i}{6}\left\{((2q^iq^j+q^2 g^{ij})\lambda_\eps-q^2 g^{ij}) [T^a,G^{ib}]G^{jb}\right.\nonumber\\
&+&3(\lambda_\eps +1)M_b^2[[T^a,G^{ib}],G^{jb}]-3 (\lambda_\eps +2)\left([[T^a,G^{ib}],[\delta \hat m,[\delta \hat m,G^{jb}]]]\right.\nonumber\\
&+&\left.\left.[[\delta \hat m,G^{ib}],[T^a,[\delta \hat m,G^{ib}]]]\right)\right\},
\eea
where in the evaluations   $p_0\to \delta m_{in}$ and $p_0+q_0\to \delta m_{out}$. 
Combining the polynomial pieces and using that $[\delta \hat m,T^a]=[\delta \hat m,\hat G^2]=[\delta \hat m,G^{ib} T^a G^{ib}]=0$ lead  to the result:
\bea
(A+B)^{\text{poly}}&=&-\frac{\lambda_\eps+1}{(4\pi)^2} \frac{\vec {q}^{\;2}}{4 F_\pi^2} \,  T^a\nonumber\\
(C+D+E)^{\text{poly}}&=&\frac{\lambda_\eps-3}{(4\pi)^2}\left( \frac{\mathring g_A}{4F_\pi} \right)^2\vec {q}^{\;2} \,T^a
\eea
As required by the AGT,   when $q\to 0$ the UV divergencies and polynomial terms vanish  for    all the $SU(3)$ vector charges of the baryon spin-favor multiplet.  The calculation of the finite non-analytic contributions has been carried out in previous work \cite{Flores-Mendieta:2014vaa}, and will not be revisited here. 

The only counter term required is the one proportional to $g_E$ in Eq.(\ref{eq:L3}), where $\beta_{g_E}=\frac{1}{(4 F_\pi)^2}(4-\mathring{g}_A^2)$, 
%{\red check this}, 
and which provides the only analytic contribution to the octet and decuplet charge radii up to the order of the calculation. More details will be presented elsewhere in a study of the form factors of the the vector currents. In the context of the charge form factors,   studies implementing the $1/N_c$ expansion for extracting the long distance charge distribution of the nucleon has been carried out in Refs. \cite{Granados:2013moa,Granados:2016jjl,Alarcon:2017asr,Alarcon:2017lhg}.
 
\section{Axial couplings}
\label{sec:Axial}
\begin{center}
\begin{figure}[h]
\centerline{
\includegraphics[width=15cm,angle=0]{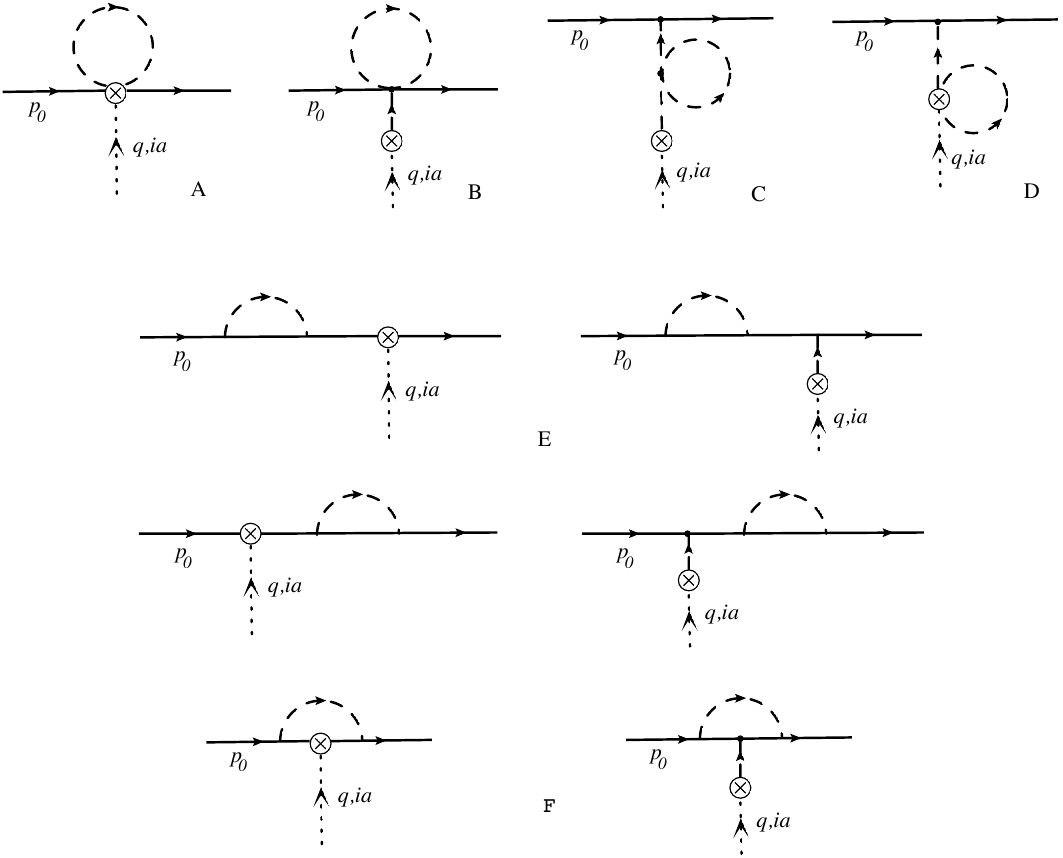}  }
\caption{Diagrams contributing to the 1-loop corrections to the axial vector currents.     }
\label{fig:1-loop-AC}
\end{figure}
\end{center}
The  axial vector currents are studied to one-loop. At the tree level the axial vector currents have two contributions, namely the contact term and the GB pole ones, and reads:
\beq
A^{\mu a}=\mathring g_A G^{ja} ( g^{\mu}_ j-\frac{q^\mu q_j}{q^2-M_a^2}).
\eeq
In the non-relativistic limit, or equivalently large $N_c$ limit, the time component of the axial vector current is suppressed with respect to the spatial components.  The couplings associated with the latter are analyzed below to  $\ord{\xi^2}$. 

At the leading order the axial couplings are all given in terms of   $\mathring g_A$. For $N_c=3$:  $F=\mathring g_A/3$, $D=\mathring g_A/2$, and the axial coupling in the decuplet baryons is ${\cal{H}}=\mathring g_A/6$.

The one-loop diagrams contributing at that order are shown in Fig. \ref{fig:1-loop-AC}.

 The matrix elements of interest for the axial currents are $\langle\B'\mid A^{ia}\mid\B\rangle$ evaluated at vanishing external 3-momentum. The axial couplings $g_A^{\B\B'}$ are conveniently defined by:
 \beq
 \langle\B'\mid A^{ia}\mid\B\rangle=g_A^{\B\B'} \; \frac{6}{5}\;{ \langle \B' \mid G^{ia} \mid \B \rangle} \, ,
\label{eq:gA-formula}
 \eeq
which  are $\ord{N_c^0}$. The $\ord{N_c}$ of the matrix elements of the axial currents is due to the operator $G^{ia}$. The factor $6/5$ mentioned earlier is included so  that $g_A^{NN}$ at $N_c=3$ exactly corresponds to the usual nucleon $g_A$, which has the value $1.267\pm0.004$ \cite{Patrignani:2016xqp}. 
 
The results for the one-loop diagrams are the following:

\bea
A&=&-g^\mu_i\frac{\mathring g_A}{2F_\pi^2} f^{abc}f^{cdb}G^{id}I(0,1,M_b)\nonumber\\
B&=&\frac{\mathring g_A}{6F_\pi^2} \frac{q^\mu q_i}{q^2-M_a^2}f^{abc}f^{cdb}G^{id}I(0,1,M_b)\nonumber\\
C&=&\frac{2\mathring g_A}{3F_\pi^2} \frac{q^\mu q_i}{q^2-M_d^2}f^{abc}f^{cdb}G^{id}I(0,1,M_b)\nonumber\\
D&=&-\frac{\mathring g_A}{3F_\pi^2} \frac{q^\mu q_i}{q^2-M_a^2}f^{abc}f^{cdb}G^{id}I(0,1,M_b)\nonumber\\
E&=&\frac 12 \mathring g_A (g^{\mu}_{ i}-\frac{q^\mu q_i}{q^2-M_a^2})\{G^{ia},\delta \hat Z_{\text{1-loop}}\}\nonumber\\
F&=&i (g^\mu_i-\frac{q^\mu q_i}{q^2-M_a^2} )\mathring g_A  \left( \frac{\mathring g_A}{F_\pi} \right)^2 \sum_{n_1,n_2} G^{jb} {\cal{P}}_{n_2} G^{ia}  {\cal{P}}_{n_1} G^{kb}\;\frac{1}{q_0-\delta m_{n_2}+\delta m_{n_1}}\nonumber\\
&\times&(H_{jk}(p_0-\delta m_{n_1},M_b)-H_{jk}(p_0+q_0-\delta m_{n_2},M_b))
\label{eq:1-loop-AC}
\eea
The corresponding polynomial terms of these one-loop contributions are:
\bea
\label{eq:acuvi}
A^{\text{poly}}&=&	\frac{1}{(4 \pi)^2}	\frac{\mathring g_A}{2F_\pi^2} (\lambda_\eps+1) g^\mu_i f^{abc}f^{bcd}G^{id} M_b^2	\nonumber\\
B^{\text{poly}}&=&	-\frac{1}{(4 \pi)^2}		\frac{\mathring g_A}{6F_\pi^2} (\lambda_\eps+1) \frac{q^\mu q_i}{q^2-M_a^2} f^{abc}f^{bcd}G^{id} M_b^2		\nonumber\\
C^{\text{poly}}&=&-\frac{1}{(4 \pi)^2}		\frac{2\mathring g_A}{3F_\pi^2} (\lambda_\eps+1) \frac{q^\mu q_i}{q^2-M_d^2} f^{abc}f^{bcd}G^{id} M_b^2				\nonumber\\
D^{\text{poly}}&=&\frac{1}{(4 \pi)^2}		\frac{\mathring g_A}{3F_\pi^2} (\lambda_\eps+1) \frac{q^\mu q_i}{q^2-M_a^2} f^{abc}f^{bcd}G^{id} M_b^2				\nonumber\\
E^{\text{poly}}&=&	\frac{1}{(4 \pi)^2}	\frac 12 \mathring g_A \left( \frac{\mathring g_A}{F_\pi} \right)^2  (g^{\mu}_{ i}-\frac{q^\mu q_i}{q^2-M_a^2})			\\
&\times& \{G^{ia},(\lambda_\eps+1)M_b^2 G^{jb} G^{jb}-2(\lambda_\eps+2) G^{jb}[\delta\hat m,[\delta \hat m,G^{jb}]]\}\nonumber\\
F^{\text{poly}}&=&
-\frac{1}{(4 \pi)^2}	
\mathring g_A 	 
\left( 
\frac{\mathring g_A}{F_\pi} 
\right)^2   (g^\mu_i-\frac{q^\mu q_i}{q^2-M_a^2} )\Big((\lambda_\eps+1)M_b^2 G^{jb} G^{ia} G^{jb}	  	\nonumber\\
&-& \frac 23  ( \lambda_\eps+2)
\left(G^{jb}G^{ia}[\delta\hat m,[\delta\hat m,G^{jb}]]+[\delta\hat m,[\delta\hat m,G^{jb}]]G^{ia}G^{jb}-[\delta\hat m,G^{jb}]G^{ia}[\delta\hat m,G^{jb}]\right)\Big).\nonumber
\eea
The conservation of the axial currents is readily checked in the chiral limit. 
At this point it is important to check the cancellation of the $N_c$ power counting violating terms shown in the polynomial terms of diagrams $E$ and $F$. Such terms cancel in the sum, as it is easy to show using the results displayed in Appendix \ref{app:opred} for the axial vector currents. One obtains:
\bea
(E+F)^{\text{poly}}&=&	\frac{1}{(4 \pi)^2}	  \mathring g_A \left( \frac{\mathring g_A}{F_\pi} \right)^2  (g^{\mu}_{ i}-\frac{q^\mu q_i}{q^2-M_a^2})	\nonumber\\
&\times&\left( (\lambda_\eps+1)\,\frac{1}{6}\, B_0\,(23\, m^0 G^{ia}+\frac{11}{4} d^{abc} m^b G^{ic}+\frac 53 m^a S^i)\right.\nonumber\\
&+&\left. (\lambda_\eps+2)\frac{C_{\rm HF}^2}{N_c^2}\left(\left(1-\frac{N_c(N_c+6)}{3}\right)G^{ia}+\frac{11}{6}(N_c+3) S^iT^a\right.\right.\nonumber\\
&-&\left.\left.\frac{8}{3}\{\hat S^2,G^{ia}\}
-\frac{4}{3} S^i\{S^j,G^{ja}\}+\frac{11}{6} \hat S^2 G^{ia}\hat S^2\right)\right)  
\label{eq:ACUV2}
\eea
 
The quark mass dependent UV divergencies  are $\ord{m_q/N_c}$, and the quark mass independent  ones  give a term proportional to $G^{ia}$, i.e., to the LO term but suppressed by a factor $1/N_c$, while the rest of the terms are $\ord{1/N_c^2}$ or higher.  The cancellation mechanism clearly requires the contributions from the wave function renormalization factors (diagrams E), and    it is rather subtle  as it requires an explicit  and lengthy calculation  starting from Eq. (\ref{eq:acuvi}).   To obtain the counter-terms   the relations given in Appendix \ref{app:opred} are used.  The counter-terms are contained in the Lagrangians $ { \cal{L}}_\B^{(1,2,3)}$, and the corresponding   $\beta$ functions are the ones shown in Table \ref{tab:betafunctionsAC}. In addition to $\mathring {g}_A$, there are seven LECs  that are necessary to renormalize the axial vector couplings for generic $N_c$. For $N_c=3$ the terms proportional to $C_{1,2,3}^A$ are linearly dependent and one can be eliminated.  At $N_c=3$, after considering isospin symmetry, there are thirty four  axial couplings associated with the axial currents mediating transitions in the spin-flavor multiplet of baryons. This means that there are twenty seven relations among those couplings that must be satisfied at the order of the present calculation. Such relations are straightforward  to derive with the results provided here, and they should eventually become  one  good test for their  LQCD calculations. It should be noted that in general the relations    dependent on $N_c$ explicitly. 

\begin{centering}
\begin{table*}[ttt]
 {
\begin{tabular}{|c|c|c|c|}
\hline\hline
LEC & $  F_\pi^2 \beta $ & LEC & $ F_\pi^2 \beta/\Lambda^2   $ \\\hline
$\mathring{g}_A$ 	& $\mathring{g}_A^3 \frac{C_{\rm HF}^2}{3} $ 		&$D_1^A$&$- \frac{1}{48}\mathring{g}_A (36+23 \mathring g_A^2)  $\\
 $C_1^A$ 		&$ -\frac{11}{6}\mathring{g}_A ^3 C_{\rm HF} ^2\frac{N_c+3}{N_c}  $ 		&$D_2^A$&$-\frac{5}{144}\mathring g_A^3$\\
  $C_2^A$		& $\frac 12 \mathring g_A^3 C_{\rm HF}^2 \frac{1-2N_c}{N_c}$			&$D_3^A(d)$&$-\frac{1}{192}\mathring{g}_A (36+11\mathring g_A^2) $\\
$C_3^A$		 	& $\frac 83 \mathring{g}_A ^3  C_{\rm HF}^2 $ 		&$D_3^A(f)$& 0\\	
 $ C_4^A$ 		&$ \frac 83 \mathring{g}_A ^3  C_{\rm HF}^2  $ 			&						&\\
\hline\hline 
\end{tabular}
}
\caption{$\beta$ functions for counter  terms contributing to the axial-vector currents. }  
\label{tab:betafunctionsAC}
\end{table*}
\end{centering}

The one-loop corrections to the axial currents are such that they do not contribute to the Goldberger-Treiman discrepancies  (GTD) \cite{Goity:1999by}. The discrepancies are given by terms in the Lagrangian of $\ord{\xi^3}$, namely:
\beq
 { \cal{L}}_\B^{(3)}=\cdots+i  \B^\dagger(g_{GTD} [\nabla^i,\tilde\chi_-^a] G^{ia}+g_{GTD}^0 \partial^i \chi_-^0 S^i)\B.
\eeq
As noted in  \cite{Goity:1999by} there are three LECs determining the spin 1/2   GTD  in $SU(3)$. The $1/N_c$ expansion shows that those LECs are actually determined by  the  two shown above,  which also determine the GTDs of the decuplet baryons. 

The following observations are important: if   the non-analytic contributions to the corrections to the axial couplings are disregarded,   the corrections $\ord{N_c}$ and $\ord{ N_c^0}$ to the matrix elements in $S=1/2$ and 3/2 baryons  due to the counter terms are as expected $\ord{p^2}$, i.e., proportional to quark masses. On the other hand the terms independent of quark masses are $\ord{1/N_c}$, i.e., spin symmetry breaking is suppressed by $\ord{1/N_c^2}$ with respect to the leading order, as it was noted long ago \cite{Dashen:1993jt}. This indicates that the effects of   spin-symmetry breaking are more suppressed than  the $SU(3)$ symmetry breaking   ones  \cite{Dai:1995zg,FloresMendieta:1998ii,FloresMendieta:2006ei}. It is important to note that at tree level NNLO the axial couplings satisfy some $N_c$ independent relations. For the case of   $\Delta Y=0$ couplings within the baryon octet and decuplet,   in the $I=1$ case the first relation below follows, and in the     $I=0$ ($\eta$ channel) case there are GMO and ES relations, namely:
\bea
\left(\frac{g_A}{g_V}\right)^{\pi \Delta}+\frac 35 \left(\frac{g_A}{g_V}\right)^{\pi\Xi^*}-\frac{8}{5} \left(\frac{g_A}{g_V}\right)^{\pi \Sigma^*}&=&0\nonumber\\
2 (g_A^{\eta N}+g_A^{\eta \Xi})-3 g_A^{\eta \Lambda}-g_A^{\eta \Lambda}&=&0   \nonumber\\
g_A^{\eta \Sigma^*}-g_A^{\eta \Delta}=g_A^{\eta \Xi^*}-g_A^{\eta  \Sigma^*}&=&g_A^{\eta \Omega}-g_A^{\eta \Xi^*}
\eea
  These relations are only violated by finite non-analytic terms. Additional relations are straightforward to derive for other couplings, such as those involving the $\Delta Y=\pm 1$ and the octet to decuplet off diagonal ones. Such relations will be a good tool to check results obtained in LQCD calculations of the axial couplings.

At LO and using $\left(\frac{g_A}{g_V}\right)^{\pi N}=1.267\pm 0.004$ for the nucleon, it follows that   $\left(\frac{g_A}{g_V}\right)^{K N\Lambda}= 0.760 $,  $\left(\frac{g_A}{g_V}\right)^{K N\Sigma}=-0.253  $, and $\left(\frac{g_A}{g_V}\right)^{K \Sigma\Xi}=\left(\frac{g_A}{g_V}\right)^{\pi N}$, to be compared  with the ones obtained from semi-leptonic hyperon decays \cite{Cabibbo:2003cu} $0.718\pm0.015$, $-0.340\pm 0.017$ and $1.32\pm 0.20$ respectively.  The NLO  $SU(3)$ breaking corrections are evidently necessary.  On the other hand, the coupling $g_A^{N\Delta}$ is at LO equal to $g_A$, while its phenomenological value extracted from the width of the $\Delta$ assuming a vanishing GTD is equal to $1.235\pm 0.011$ \cite{CalleCordon:2012xz,Cordon:2013era}, which shows a remarkably small breaking of the spin-symmetry.  This seems to be  in line with what was discussed above, namely that spin symmetry breaking is suppressed with respect to $SU(3)$ breaking by one extra order in $1/N_c$.  In the following subsections the results for the axial couplings are confronted with recent LQCD calculations.

\subsection{Fits to LQCD Results}
While LQCD calculations of the axial coupling of  the nucleon  have a long history,   calculations involving hyperons  and including the decuplet baryons are very recent. Indeed, the first such calculations were carried out by C. Alexandrou et al \cite{Alexandrou:2016xok}, where the   axial couplings associated with the two neutral $\Delta {\cal{S}}=0$ currents for transitions within the octet and within the decuplet baryons  were obtained.  They used a twisted mass Wilson action adapted to 2+1+1 flavors (the calculation includes charmed baryons). The results in  \cite{Alexandrou:2016xok} show the a similar recurring issue in LQCD calculations of the nucleon's axial coupling,  which  turn out to be from  5 to 10 \% smaller than the physical value. Recent calculations of $g_A^N$ have been able to give  consistent results \cite{Berkowitz:2017gql}, but those  calculations are still missing for  hyperons and the baryon decuplet.

In this subsection the results  \cite{Alexandrou:2016xok},  are fitted with the effective theory.  The LECs that can be fitted with these results are: $\mathring g_A, ~\delta \mathring g_A$ (which is a $1/N_c$ correction to $\mathring g_A$ and needed for a counterterm), and $C_{1,3}^A,  ~D_{1,2,3}^A$.  
Using the definition of couplings in Eq. \ref{eq:gA-formula},
the results shown above for the UV divergencies of the one-loop contributions imply that: $\delta g_A^{aBB'}(UV div)/g_A^{aBB'}=\ord{C_{\rm HF}/N_c}+\ord{m_q/N_c}$.
At   LO,  $g_A^{aBB'}=g_A^{N}=1.267\pm0.004$. The relations  between the couplings $g_A^{aBB'}$ and the ones  displayed in  \cite{Alexandrou:2016xok} are the following:\bea
\langle B_8\mid A^{i=0\;3}\mid B_8\rangle&=& \frac{1}{2} g_A^{B_8}\nonumber\\
\langle B_{10}\mid A^{i=0\;3}\mid B_{10}\rangle&=& \frac{1}{6} g_A^{B_{10}}\nonumber\\
\langle B_8\mid A^{i=0\;8}\mid B_8\rangle&=& \frac{1}{2\sqrt{3}} g_8^{B_8}\nonumber\\
\langle B_{10}\mid A^{i=0\;8}\mid B_{10}\rangle&=& \frac{1}{6\sqrt{3}} g_8^{B_{10}},
\eea
where $B_{8,10}$ is an octet (decuplet) baryon with spin projection $+1/2$, and the couplings on the RHS are those used in  \cite{Alexandrou:2016xok}  and displayed in Tables IV and V of that reference. The LQCD results are given for several values of $M_\pi$ by keeping $m_s$ approximately fixed. The values of $M_\pi$ for the different cases are given in Table I of  \cite{Alexandrou:2016xok}, and the corresponding $M_K$ is determined  using the physical masses by the LO relation: $M_K^2={M_K}_{\rm phys}^2+\frac{1}{2}(M_\pi^2-{M_\pi}_{\rm phys}^2)$, which corresponds to keeping $m_s$ fixed. 
While for general $N_c$   the nine terms associated with the LECs in Table \ref{tab:betafunctionsAC} are linearly independent, at $N_c=3$ the term associated with $C_2^A$ becomes linearly dependent with the
LO term, and thus its effects are absorbed into $\delta \mathring g_A$. In the case of the LQCD results being fitted here there is an additional linear dependency, namely that of the term $C_4^A$ which becomes linearly dependent with the term $C_3^A$ . So the fit will involve
seven NLO LECs in addition to $\mathring g_A$.
The results of the fits are shown in Table \ref{tab:AVCfits}.
\begin{table}[h!]\small
    \begin{tabular}{cccccccccccc}
        \hline      \hline  Fit &$\chi^2_{\rm dof}$  & $\mathring g_A$ & $\delta\mathring g_A$&$C_1^A$ &$C_2^A$ &$C_3^A$ &$C_4^A$ & $D_1^A$ & $D_2^A$ & $D_3^A$ & $D_4^A$
        \\ \hline
        LO & 3.9 &1.35 &-  &-  & - & - & -  & -  & -&-  &- \\
        NLO Tree& 0.91 &1.42&-   &-0.18  & - &-  &-&  - &  0.009 & - &-  \\
           NLO Full &1.08& 1.02& 0.15& -1.11& 0.& 1.08& 0.& -0.56& -0.02& -0.08& 0.\\
&1.13& 1.04& 0.08& -1.17& 0.& 1.15& 0.& -0.59& -0.02& -0.09& 0.\\ 
&1.19& 1.06& 0.& -1.23& 0.& 1.21& 0.& -0.62& -0.03& -0.09& 0.\\
\hline\hline
    \end{tabular}
    \caption{LECs obtained by fitting to the LQCD results presented in Tables IV and V of Ref. \cite{Alexandrou:2016xok}. The results  correspond to making  the choices   $\Lambda=\mu=m_\rho$ .   In the NLO full fits $C_{\rm HF}=250$ MeV, and  $\mathring g_A$ is given as input, displaying fits for three different   values.  } 
    \label{tab:AVCfits}
\end{table}   
The LO fit, which involves only fitting the LO value of $\mathring g_A$, shows a remarkably  good approximation to the full set of the LQCD results. This is clearly aided by the very small dependency on $M_\pi$  of the LQCD results. It also shows the very good approximate spin-flavor symmetry that relates axial couplings in the octet and decuplet. The LO fit implies that  $g_A^N=1.13$ for the physical pion mass.  A fit where only tree contributions are included up to the NNLO gives a very precise description of the LQCD results. Indeed, turning off some of the LECs as indicated in Table \ref{tab:AVCfits} provides a consistent fit, and corresponds in this case to $g_A^N=1.15$. Note that in this case $\delta\mathring g_A$, which is required to cancel an UV divergency proportional to the leading term, can be turned off, as it is only required when the loop contributions are included.

 The full NLO fit is more complicated. 
 Although the implemented consistency with the $1/N_c$ expansion gives an important reduction of the non-analytic contributions, these are still significant. The most significant issue in this case becomes the determination of the LO $\mathring g_A$. If it is used as a fitting parameter, then the fit naturally drives it down to small values,   suppressing  the non-analytic contributions. Such a situation is unrealistic, and therefore an strategy is needed.   The  problem originates  in the  need to renormalize $\mathring g_A$, as there is an UV divergency proportional   to the LO term of the axial current. This is performed using $\delta\mathring g_A$, which is suppressed by one power in $1/N_c$ with respect to $\mathring g_A$. Fixing both the LO $\mathring g_A$  and the counter-term would thus require information at different values of $N_c$, which is not accessible at present. One possible approach is to fix $\mathring g_A$ to the value obtained with the LO fit, and then fit the higher order LECs. This however fails because the resulting fit has too large a $\chi^2$.  Another strategy is to input several different values of  $\mathring g_A$, and determine an approximate range for it based of obtaining a $\chi^2$ that is acceptable. Finally   a different strategy can be used   involving  additional observables: for instance, as mentioned earlier,    the value for $\mathring g_A$ could be obtained by matching to $\Delta_{\rm GMO}$,   giving a value for $\mathring g_A/F_\pi$, which in $\Delta_{GMO}$ should be taken at LO. In that case, and  in the physical case one obtains  $\mathring g_A\sim 1.15$  when $F_\pi=93$ MeV.  This however  cannot be used for  the present LQCD results, because they have the mentioned issue of  extrapolating  to too low of a value for $g_A^N$ at the physical point. In that case a  correspondingly smaller value should be used, namely $\mathring g_A\sim 1.05$ or so. The NLO fit with such an input for  $\mathring g_A$ is almost consistent, and is shown in Table \ref{tab:AVCfits} for three different input values.  The extrapolation of those fits to the physical $M_\pi$ give a rather low value, $g_A^N\sim 0.97$. This value is increased if only the $LQCD$ results in  \cite{Alexandrou:2016xok} for the nucleon 
 are included, namely  $g_A^N\sim 1.05$. The effective theory is also checked to fit the most recent results on $g_A^N$ \cite{Berkowitz:2017gql}, where the LQCD result agreees with the physical value.  Clearly, it is necessary to await additional lattice calculations of the octet and decuplet axial couplings in order to have a thorough  test of  the effective theory vis-\'a-vis LQCD.

  Ultimately, in order to have the LECs in BChPT$\times 1/N_c$ fully determined,  a global analysis  involving  LQCD calculations of a complete set of observables is necessary. This requires   the LQCD determination of the quark mass dependencies of the observables, and also the possibility of results for different values of $N_c$, which is a more difficult task, but which has already been initiated with the baryon masses for two flavors \cite{DeGrand:2012hd}, and   which  has been analyzed with the effective theory  \cite{Cordon:2014sda}.

\section{Summary}
\label{sec:Conclusions}

Chiral symmetry and the   expansion in $1/N_c$   are two fundamental aspects of QCD. The former is known to play a crucial role in light hadrons, and there are multiple indications that the latter is also important,    in particular for baryons.  In the context of effective theories, it is therefore crucial to incorporate   those two aspects of   QCD  consistently. 
This is possible with the combined  Chiral and $1/N_c$   expansions. In the present work that framework for baryons in $SU(3)$ was implemented using the $\xi$-expansion. The renormalization to one-loop for baryon masses and currents were presented for generic $N_c$, and   LQCD results for masses and axial couplings were analyzed. This work serves as a basis for further applications, where it is expected that the  improved convergence of the effective theory    will have a significant  impact, which should be particularly important  in the case of three flavors. 

In the case of three flavors, there are numerous parameter free relations   that hold at tree level NNLO in the $\xi$ expansion, such as GMO, ES, and various other relations for $\sigma$ terms and axial couplings. Those relations have calculable corrections    given solely by the non-analytic loop contributions, thus providing useful  tests  for the accuracy of the effective theory and also   serving as control tests  of LQCD results through those same relations. 

It is important to emphasize  the  importance of the decuplet in the effective theory, which has a key role in taming the non-analytic contributions and thus improving the convergence, as it is clearly manifested in particular in the axial couplings.   This   improvement in the behavior of the effective theory when it is made consistent with the $1/N_c$ expansion permeates other observables, such as the mass relations and vector charges, as well as  virtually any other observable,  such as in pion-nucleon scattering, in Compton scattering, etc.

\begin{acknowledgments}
The authors thank  Rub\'en Flores Mendieta for useful discussions and comments on the manuscript,   Christian Weiss  and  Jos\'e Manuel  Alarc\'on for useful discussions,    and Dina Alexandrou for  communications on LQCD results for the axial couplings.
This work was supported by DOE Contract No. DE-AC05-06OR23177 under which JSA operates the Thomas Jefferson National Accelerator Facility, and by the National Science Foundation  through grants PHY-1307413 and PHY-1613951. 
\end{acknowledgments}
\newpage

\appendix

\section{Spin-flavor algebra and operator bases }
\label{app:Algebra}

The $4N_f^2-1$ generators of the spin-flavor group $SU(2 N_f)$  consist of the three spin generators $S^i$, the $N_f^2-1$   flavor $SU(N_f)$ generators  $T^a$, and the remaining $3(N_f^2-1)$   spin-flavor generators $G^{ia}$. The commutation relations are:
\bea
~& [ S^i,S^j ]= i \eps^{ijk}S^k,~~ [ T^a,T^b ]=i f^{abc} T^c ,~~ [ T^a,S^i ]=0& ,\nonumber\\    
~& [ S^i,G^{ja} ]=i  \eps^{ijk} G^{ka},~~ [T^a,G^{ib}]=i f^{abc} G^{ic}&,\nonumber\\   
~ & [ G^{ia},G^{jb} ]= \frac{i}{4}\delta^{ij}f^{abc}T^c+\frac{i}{2 N_f}\delta^{ab} \eps^{ijk} S^k+\frac{i}{2}\eps^{ijk}d^{abc}G^{kc}&.
\label{eq:commutation-relations}
\eea
 
In representations with $N_c$ indices (baryons), the generators $G^{ia}$ have matrix elements $\ord{N_c}$ on states with $S=\ord{N_c^0}$. A contracted  $SU(6)$ algebra  is  defined by the generators $\{S^i,I^a,X^{ia}\}$, where $X^{ia}=G^{ia}/N_c$.  In large $N_c$, the generators $X^{ia}$ become semiclassical as $[X^{ia},X^{jb}]=\ord{1/N_c}$, and have matrix elements $\ord{1}$ between baryons.

The  symmetric irreducible representation of $SU(6)$ with $N_c$ Young boxes decomposes into the following $SU(2)_{\text{spin}}\times SU(3)$ irreducible representations: $[S,(p,q)]=[S,(2S,\frac{1}{2}(N_c-2S))]$, $S=1/2,\cdots,N_c/2$ (assumed $N_c$ is odd). The baryon states are then denoted by: $\mid \! S S_3,YII_3\rangle$. Clearly the spin $S$ of the baryons determines its $SU(3)$ irreducible representation.

Some useful details about the contents of $SU(3)$ multiplets are in order.
For a given irreducible representation  $(p,q)$, the range of hypercharge is:
\bea
Y_{\text{min}}(p,q)=-\frac{2p+q}{3}\leq Y\leq Y_{\text{max}}(p,q)=\frac{p+2q}{3}
\eea
Defining:
\bea
\bar{Y}(p,q)&=& Y_{\text{max}}(p,q)-q\nonumber\\
\bar{Y}'(p,q)&=&Y_{\text{min}}(p,q)+q,
\eea
where $\bar{Y}>\bar{Y}'$ if $p>q$, and viceversa.
The possible  isospin values for a given $Y$ are as follows:
\bea
\text{if~~~} p\geq q:~~I(Y)&=&\left\{\begin{array}{lr}
	\text{if~~~}    Y\geq \bar{Y}      :&   \frac{1}{2}(p-Y_{\text{max}}+Y),\cdots,\frac{1}{2} (p+Y_{\text{max}}-Y)    \\	\text{if~~~}  \bar{Y}'\leq Y<\bar{Y}        :&    \frac{1}{2} (p-Y_{\text{max}}+Y),\cdots,\frac{1}{2} ( p+Y_{\text{max}}+Y-2\bar Y)   \\	\text{if~~~}  Y_{\text{min}}\leq Y<\bar{Y}'        :&  \frac{1}{2} (q+Y_{\text{min}}-Y),\cdots,\frac{1}{2} ( q+Y-Y_{\text{min}} )       
	\end{array}\right.\nonumber\\
	\text{if~~~} q\geq p:~~I(Y)&=&\left\{\begin{array}{lr}
		\text{if~~~}    Y\geq \bar{Y}'      :&   \frac{1}{2}(p-Y_{\text{max}}+Y),\cdots,\frac{1}{2} (p+Y_{\text{max}}-Y)    \\	\text{if~~~}  \bar{Y}\leq Y<\bar{Y}'        :&    \frac{1}{2} (p+2 \bar{Y}'-Y_{\text{max}}-Y),\cdots,\frac{1}{2} ( p+Y_{\text{max}}- Y)   \\	\text{if~~~}  Y_{\text{min}}\leq Y<\bar{Y}        :&  \frac{1}{2} (q+Y_{\text{min}}-Y),\cdots,\frac{1}{2} ( q+Y-Y_{\text{min}} )       
	\end{array}\right.\nonumber
\eea 

\subsection{Matrix elements of spin-flavor generators} 
The $SU(6)$ algebra involved in the calculations is quite lengthy and laborious, and therefore it is useful to provide basic details that are of help in implementing it. Here the matrix elements of the $SU(6)$ generators are given;  additional details can be found in \cite{Matagne:2014lla}.
In general the matrix elements of a $SU(2)_{\text{spin}}\times SU(3)$ tensor operator between baryons of the form $\mid \! SS_3,R\;YII_3\rangle$, where $R$ is the irreducible representation of $SU(3)$ to which the state belongs, will be given according to the Wigner-Eckart theorem in terms of reduced matrix elements and Clebsch-Gordan coefficients as follows:
\bea
\langle S'S'_3,R'\;Y'I'I'_3\mid O^{\ell \ell_3}_{\tilde R \tilde Y\tilde I\tilde I_3}\mid \! SS_3,R\;YII_3\rangle &=&\frac{1}{\sqrt{2S'+1}\sqrt{\text{dim}R'}}\;\langle SS_3,\ell\ell_3\mid \! S'S'_3\rangle \\
	\times  \sum_{\gamma } \langle S',R'\mid\mid O^{\ell }_{\tilde R  }\mid\mid \! S,R\rangle_\gamma&&\!\!\!\!
	\left\langle
	\begin{array}{cc}
		R  & \tilde R       \\
		Y ~I ~I_{3}     & \tilde Y~\tilde I~ \tilde I_3
	\end{array}
	\right|   \left.
	\begin{array}{c}
		 R'          \\
Y' ~ I' ~ I'_3
	\end{array} \right\rangle_\gamma  ~~,\nonumber
\eea
where $\gamma$ indicates the   re-coupling index  in $SU(3)$ for $R\otimes \tilde R\to R'$. Matrix elements of the spin-flavor generators between baryon states in the spin-flavor symmetric representation are then given by:
\bea
\langle S'S'_3,Y'I'I'_3\mid \! S^m\mid \! SS_3,YII_3\rangle&=& \delta_{SS'}\delta_{YY'}\delta_{II'}\delta_{I_3I'_3}\sqrt{S(S+1)} \langle SS_3,1m\mid \! S'S'_3\rangle\nonumber\\
\langle S'S'_3,Y'I'I'_3\mid T^{yii_3}\mid \! SS_3,YII_3\rangle&=&\delta_{SS'}\delta_{S_3S'_3}\frac{1}{\sqrt{\text{dim}(2S,\frac 12(N_c-2S))}}\langle S\mid\mid T\mid\mid \! S\rangle\nonumber\\
&\times& 	\left\langle
\begin{array}{cc}
	(2S,\frac 12(N_c-2S))  & (1,1)      \\
	Y ~I ~I_{3}     & yii_3
\end{array}
\right|   \left.
\begin{array}{c}
(2S,\frac 12(N_c-2S))          \\
	Y' ~ I' ~ I'_3
\end{array} \right\rangle_{\gamma=1}\nonumber\\
\langle S'S'_3,Y'I'I'_3\mid G^{m,yii_3}\mid \! SS_3,YII_3\rangle&=&
\frac{\langle SS_3,1m\mid \! S'S'_3\rangle}{\sqrt{2S'+1}\sqrt{\text{dim}(2S,\frac 12(N_c-2S))}}\\
\times \sum_{\gamma=1,2 } \langle S'\mid\mid G\mid\mid \! S\rangle_\gamma &&\!\!\!\!	\left\langle
\begin{array}{cc}
	(2S,\frac 12(N_c-2S))  & (1,1)      \\
	Y ~I ~I_{3}     & yii_3
\end{array}
\right|   \left.
\begin{array}{c}
	(2S,\frac 12(N_c-2S))          \\
	Y' ~ I' ~ I'_3
\end{array} \right\rangle_{\gamma}\nonumber
\eea
where the reduced matrix elements are (here $p=2S$, $q=\frac 12(N_c-2S)$):
\small{
\bea
\!\!\!\!\!\!
\langle S\mid\mid T\mid\mid \! S\rangle&=&\text{sign}(q-0^+)\sqrt{\text{dim}(	p,q) C_2(	p,q)}\nonumber\\
=\text{sign}(N_c-2S-0^+)&&\!\!\!\!\!\frac{\sqrt{(2 S+1) ({N_c}-2
		S+2) ({N_c}+2 S+4)
		({N_c} ({N_c}+6)+12 S
		(S+1))}}{4 \sqrt{6}}\nonumber\\
\langle S'\mid\mid G\mid\mid \! S\rangle_{\gamma=1}\!\!&=&\!\!\left\{\begin{array}{ll}
\text{if~~}S=S'+1:&-\frac{\sqrt{\left(4 S^2-1\right)
		\left(({N_c}+2)^2-4
		S^2\right)
		\left(({N_c}+4)^2-4
		S^2\right)}}{8 \sqrt{2}}\\
\text{if~~}S=S'-1:&-\frac{\sqrt{(4 S (S+2)+3)
		({N_c}-2 S) ({N_c}-2
		S+2) ({N_c}+2 S+4)
		({N_c}+2 S+6)}}{8 \sqrt{2}}\\
\text{if~~}S=S':&\text{sign}(N_c-2S-0^+)\frac{({N_c}+3) (2 S+1)
	\sqrt{S (S+1) ({N_c}-2 S+2)
		({N_c}+2 S+4)}}{ 
	\sqrt{6{N_c}
		({N_c}+6)+12 S (S+1)}}
 \end{array} \right. \\
\langle S'\mid\mid G\mid\mid \! S\rangle_{\gamma=2}&=&{\footnotesize -\delta_{SS'}\frac{(2 S+1) \sqrt{({N_c}-2
		S) ({N_c}+2 S+6)
		\left(({N_c}+2)^2-4
		S^2\right)
		\left(({N_c}+4)^2-4
		S^2\right)}}{8 \sqrt{2}
	\sqrt{{N_c}
		({N_c}+6)+12 S (S+1)}}}  \nonumber
\eea}
In the case of the generators $G$, $\gamma=1~ (2)$ correspond to the re-couplings $R\otimes \bar R'\to 8 ~(8')$ respectively.

\subsection{Bases of   spin-flavor  composite operators}
\label{sec:OperatorBases}
Here the bases of 2- and 3-body spin-flavor operators along with important operator relations relevant to this work are given.

There are operator relations which are valid for matrix elements in the symmetric irreducible representation of $SU(6)$.  The first ones are relations for 2-body operators \cite{Dashen:1994qi},  and are shown in Table \ref{DJMRel}.
\begin{table}[h!]
\begin{tabular}{c|c}\hline\hline
	Relation &  $SU(2)_{\text{spin}}\times SU(3)$  
	\\ \hline
	$2\,\hat S^2+3\,\hat T^2+12\,\hat G^2=\frac 52N_c(N_c+6)$ & $(\ell=0,\bf{1})$\\
	$   d^{abc}\{G^{ia},G^{ib}\}+\frac 23\{S^i,G^{ic}\}+\frac 14 d^{abc}\{T^a,T^b\}=\frac{2}{3}(N_c+3)T^c $  & $  (0,\bf{8} )$\\
	$  \{T^a,G^{ia}\}=  \frac{2}{3}(N_c+3) S^i$  & $  (1,\bf{1}) $\\
	$  \frac 13\{S^i,T^a\}+d^{abc}\{T^b,G^{ic}\}-\eps^{ijk}f^{abc}\{G^{jb},G^{kc}\} =\frac{4}{3}(N_c+3) G^{ia} $  & $   (1,\bf{8})$\\
	$-12\,\hat G^2+27\,\hat T^2-32\,\hat S^2=0    $  & $(0,\bf{1})   $\\
	$  d^{abc}\{G^{ib},G^{ic}\}+\frac 94 d^{abc}\{T^b,T^c\}-\frac{10}{3} \{S^i,G^{ia}\}   $  & $(0,\bf{8})   $\\
	$  4 \{G^{ia},G^{ib}\}^{\bf{27}}=\{T^a,T^b\}^{\bf{27}}  $  & $(0,\bf{27})  $\\
	$  d^{abc}\{T^b,G^{ic}\}=\frac 13(\{S^i,T^a\}-\eps^{ijk}f^{abc}\{G^{jb},G^{kc}\})  $  & $  (1,\bf{8}) $\\
	$  \eps^{ijk}\{G^{ja},G^{kb}\}^{\bf{10+\bar{10}} }=(f^{acd}d^{bce}\{T^d,G^{ie}\} )^{\bf{10+\bar{10}}}$  & $ (1,{\bf{10+\bar{10}} } )$\\
	$  \{G^{ia},G^{ja}\}^{\ell=2}=\frac 13 \{S^i,S^j\}^{\ell=2}  $  & $(2,\bf{1})   $\\
	$  d^{abc}\{G^{ia},G^{jb}\}^{\ell=2}=\frac 13  \{S^i,G^{ja}\}^{\ell=2} $  & $(2,\bf{8})   $\\
	\hline\hline
	\end{tabular}
	\caption{2-body identities for the $SU(6)$ generators acting on the irreducible representation $(N_c,0,0,0,0,0)$.}
	\label{DJMRel}
\end{table}
The relations in Ref.  \cite{Dashen:1994qi} are for general $N_f$, and the correspondence for $N_f=3$ given here is as follows (left Ref. \cite{Dashen:1994qi}, right  Table (\ref{DJMRel})): $0\to {\bf 1}$, $\bar s s\to{\bf 27}$, $\bar a s+\bar s a\to{\bf 10+\bar{10}}$, while there is no term $\bar a a$ for $N_f=3$.

The following identities follow from  Table (\ref{DJMRel}), namely
from the $ (0,\bf{1})$  relations:
\bea
\hat G^2&=& \frac{1}{4}\left(\frac {3}{4} N_c(N_c+6) -\frac{5}{3} S^2\right)\nonumber\\
\hat T^2&=&\frac{1}{4} \left(\frac{  {N_c} ({N_c+6})}{3}+4 \hat S^2\right),
\eea
from the $(0,\bf{8})$ relations:
\bea
d^{abc}\{G^{ib},G^{ic}\}&=&\frac 34 ( N_c+3) T^a-\frac{7}{6} \{S^i,G^{ia}\}\nonumber\\
d^{abc}\{T^{b},T^{c}\}&=&-\frac{( N_c+3)}{3}T^a+2 \{S^i,G^{ia}\},
\eea
and from the $(1,\bf{8})$ relations:
\bea
\eps^{ijk}f^{abc} \{G^{ia},G^{jb}\}&=& ( S^k T^c-(N_c+3) G^{kc})\nonumber\\
d^{abc} \{T^a, G^{ib}\}&=&2d^{abc} T^a G^{ib}=\frac{1}{3}(S^iT^c+(N_c+3) G^{ic})\nonumber\\
f^{abc}\{T^b,G^{ic}\}&=&\eps^{ijk}\{S^j,G^{ka}\},
\eea
while the rest of the identities are explicit in   Table \ref{DJMRel}.
Making use of these relations, the basis of 2-body operators can be chosen to be as shown in Table \ref{2bodybasis}:
\begin{table}[h!]
\begin{tabular}{c|c}\hline\hline
2-body operator & $(\ell,\bf{R})$\\
\hline
$\hat S^2$ & $(0,\bf{1})$\\

$ \{S^i,S^j\}^{\ell=2}$ & $( 2,\bf{1})$\\
$ \{S^i,T^a\}$ & $(1 ,\bf{8})$\\
$\{S^i,G^{ia}\}$ & $(0 ,\bf{8})$\\
$ \eps^{ijk}\{S^j,G^{ka}\}$ & $(1 ,\bf{8})$\\
$\{S^i,G^{ja}\}^{\ell=2} $ & $(2 ,\bf{8})$\\
$\{T^a,G^{ib}\}^{\bf{10+\bar{10}}} $ & $(1 ,{\bf{10+\bar{10}}})$\\
$\{T^a,T^b\}^{\bf{27}} $ & $(0 ,\bf{27})$\\
$ \{G^{ia},G^{jb}\}^{(2,\bf{27})}$ & $(2 ,\bf{27})$\\
$ \{T^a,G^{ib}\}^{\bf {27}}$ & $( 1,\bf{27})$\\
\hline\hline
	\end{tabular}
	\caption{2-body basis operators.}
	\label{2bodybasis}
\end{table}

Making use of the basis of 2-body operators, some lengthy work leads to building the basis of 3-body operators with $\ell=0,1$. That basis is displayed in Table \ref{3bodybasis}:
\begin{table}[h!]
\begin{tabular}{c|c}\hline\hline
3-body operator & $(\ell,\bf{R})$\\
\hline
$ T^a \hat S^2$ & $(0,\bf{8})$\\
$\{T^a,\{S^i,G^{ib}\}\}^{\bf{10+\bar{10}}} $ & $(0,{\bf{10+\bar{10}}})$\\
$\{T^a,\{S^i,G^{ib}\}\}^{\bf{27}} $ & $(0,{\bf{27}})$\\
$ S^i \hat S^2$ & $(1,\bf{1})$\\
$\{T^a,\{T^b,T^c\}^{\bf{27}}\} $ & $(0,\bf{8\otimes 27})$\\
$ S^i\{T^a,T^b\}^{\bf{27}}$ & $(1,\bf{27})$\\
$\{S^j,\{G^{ia},G^{jb}\}^{(2,\bf{27})}\} $ & $(1,\bf{27})$\\
$\{\hat S^2,G^{ia}\} $ & $(1,\bf{8})$\\
$\eps^{ijk}\{S^j,\{T^a,G^{kb}\}\}^{\bf{10+\bar{10}}} $ & $(1,{\bf{10+\bar{10}}})$\\$\eps^{ijk}\{S^j,\{T^a,G^{kb}\}\}^{\bf{27}} $ & $(1,{\bf{27}})$\\
$\{G^{ia},\{T^b,T^c\}^{\bf {27}}\} $ & $(1,\bf{8\otimes 27})$\\
$\{G^{ia},\{S^j,G^{jb}\}\} $ & $(1,\bf{8\otimes 8})$\\
\hline\hline
	\end{tabular}
	\caption{Operators of interest in the 3-body basis   up to $\ell=1$. }
	\label{3bodybasis}
\end{table}

\section{Building blocks for the effective Lagrangians}
\label{app:buildblocks}

In the symmetric representations of $SU(6)$  the baryon spin-flavor multiplet consists of the baryon states in the $SU(3)$ irreducible representations $(p=2S,q=\frac 12(N_c-2S))$, where $S$ is the baryon spin.  This permits a straightforward implementation of the non-linear realization of chiral $SU_L(3)\times SU_R(3)$ on the spin-flavor multiplet. The baryon spin-flavor multiplet is given by the field $\B$, where the components of the field have well defined spin, and therefore also are in irreducible representations of $SU(3)$.

 Defining as usual the Goldstone Boson fields $\pi^a$, $a=1,\cdots,8$, through the unitary parametrization $u=\exp(i\frac{\pi^a T^a}{F_\pi})$ (note that in the fundamental representation $T^a=\lambda^a/2$, with $\lambda^a$ the Gell-Mann matrices), for any isospin representation one defines a non-linear realization of chiral symmetry according to \cite{Coleman:1969sm,Callan:1969sn}:
\beq
(L,R):u=u'=R \,u\, h^\dag\!(L,R,u)=h(L,R,u)\,u\,L^\dag,
\label{eq:transformation-u}
\eeq
where $(L,R)$ is a   $SU_L(3)\times SU_R(3)$ transformation. This equation  defines $h$, and  since $h$ is a  $SU(3)$ transformation itself, it can be written as $h=\exp(i c^a T^a)$.  The chiral  transformation on the baryon multiplet $\B$ is then given by:
\beq
(L,R):\B=\B'=h(L,R,u)\B.
\label{eq:transformation-B}
\eeq
On the other hand, spin-flavor transformations of interest are the contracted ones, namely those generated by $\{S^i,I^a,X^{ia}=\frac{1}{N_c}G^{ia}\}$.
While the isospin transformations act on the pion fields in the usual way, and the spin transformations must be performed along with the corresponding spatial rotations.   The transformations generated by $X^{ia}$ are   defined  to only act on the baryons.

The effective baryon Lagrangian can be expressed in the usual way as a series of terms which are $SU_L(3)\times SU_R(3)$ invariant (upon introduction of appropriate sources; see for instance \cite{Scherer:2002tk} for details). 
The fields in the effective Lagrangian are the Goldstone Bosons parametrized by the unitary $SU(3)$ matrix field $u$ and the baryons   given by  the symmetric $SU(6)$ multiplet  $\B$.  

The building blocks for the effective theory consist of low energy operators composed in terms of the GB fields, derivatives and sources (chiral tensors), and spin-flavor composite operators (spin-flavor tensors). 

The low energy operators are the usual ones, namely:
\bea
 D_\mu&=&\partial_\mu-i \Gamma_\mu,~~~\Gamma_\mu=\Gamma_\mu^\dag=\frac{1}{2}(u^\dag(i\partial_\mu+r_\mu)u+u(i\partial_\mu+\ell_\mu)u^\dag),\nonumber\\
u_\mu&=&u^\dag_\mu=u^\dag(i\partial_\mu+r_\mu)u	-u(i\partial_\mu+\ell_\mu)u^\dag,	\nonumber\\	
\chi&=&2 B_0(s+i p)	,~~
\chi_{\pm}=u^\dag  \chi u^\dag \pm u\chi^\dag  u,\nonumber\\
F^{\mu\nu}_{L}&=&\partial^\mu\ell^\nu-\partial^\nu\ell^\mu-i[\ell^\mu,\ell^\nu],~~F^{\mu\nu}_{R}=\partial^\mu r^\nu-\partial^\nu r^\mu-i[r^\mu,r^\nu],
\label{eq:chiral-building-blocks}
\eea
where $D_\mu$ is the chiral covariant derivative,  $s$ and $p$ are scalar and pseudo-scalar sources,  and $\ell_\mu$ and $r_\mu$ are gauge sources. It is convenient to define the $SU(3)$ singlet and octet components of $\chi^\pm$ using the fundamental $SU(3)$ irreducible representation, namely:
\bea
\chi_\pm^0&=&\frac{1}{3} \langle \chi_\pm\rangle\nonumber\\
\tilde\chi_\pm&=&\chi_\pm-\chi_\pm^0=\tilde\chi_\pm^a \frac{\lambda^a}{2}
\eea
Displaying explicitly the quark masses,
\bea
\chi_{+}&=&4 B_0  {\cal{M}}_q+\cdots .
\eea
The three quark mass combinations, namely $SU(3)$ singlet, isosinglet, and isotriplet are respectively defined  to be:
 \beq
 m^0= \frac{1}{3}(m_u+m_d+m_s),~~~~~~ m^8= \frac{1}{\sqrt{3}}(m_u+m_d-2m_s),~~~~~~m^3\equiv (m_u-m_d).
 \eeq
 
The spin-flavor operators were discussed in Appendix \ref{app:Algebra}.

The leading order  equations of motion are used in the construction of the higher order terms in the Lagrangian, namely, $iD_0\B= (\frac{C_{\rm HF}}{N_c} S(S+1)+\frac{c_1}{2\Lambda}    \hat\chi_+)\B$, and $\nabla_\mu u^\mu=\frac{i}{2}\chi_-$.

\subsection*{Interaction vertices and currents at LO}
\begin{center}
\begin{figure}[h!!]
\centerline{\includegraphics[width=10.cm,angle=-0]{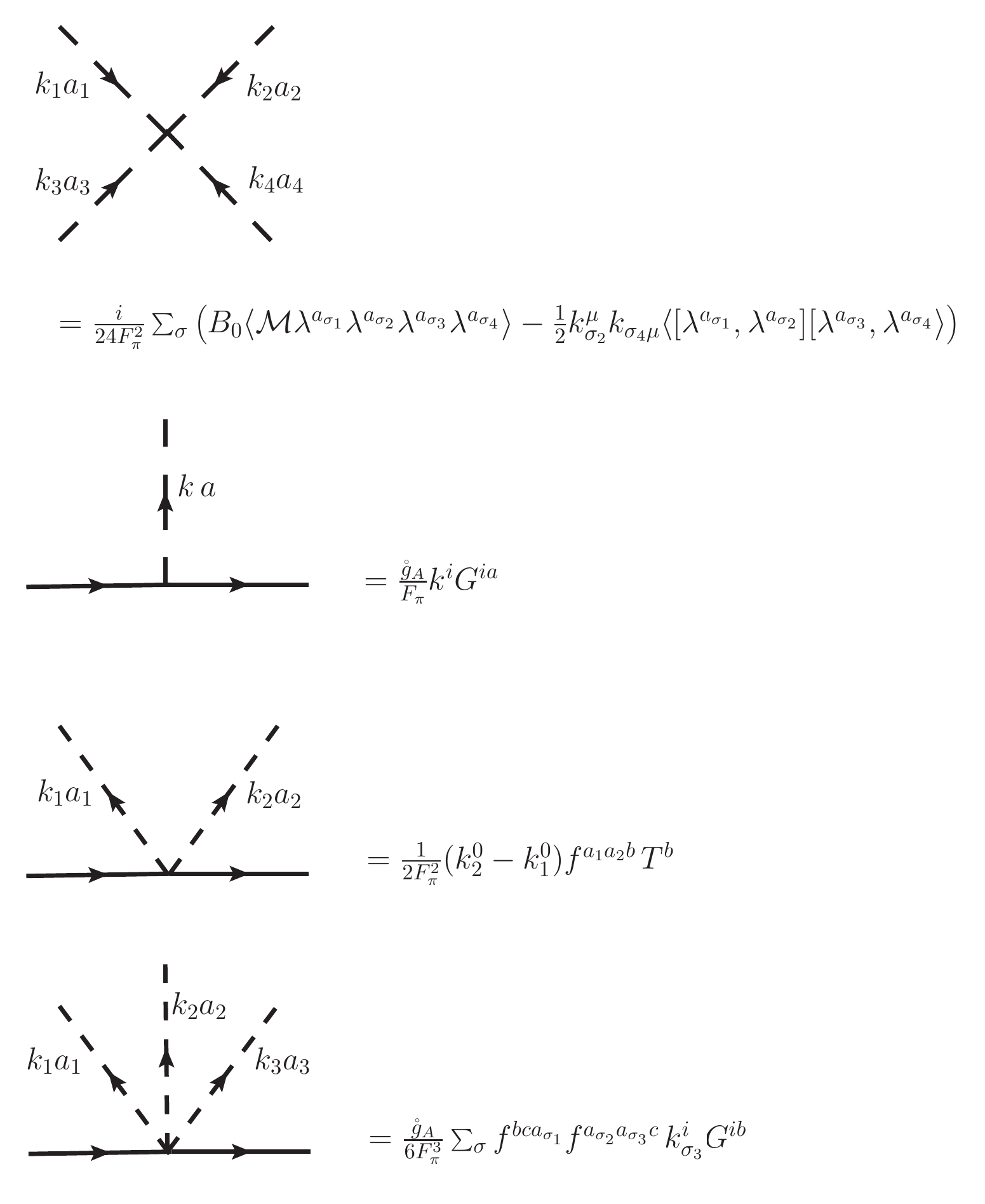}}
\caption{Interaction vertices from the LO Lagrangians. ${\cal{M}}$ is the quark mass matrix. $\sum_\sigma$ indicates sum over the corresponding permutations.}
\label{fig:vertices}
\end{figure}
\end{center}

\begin{center}
\begin{figure}[h]
\centerline{\includegraphics[width=10.cm,angle=-0]{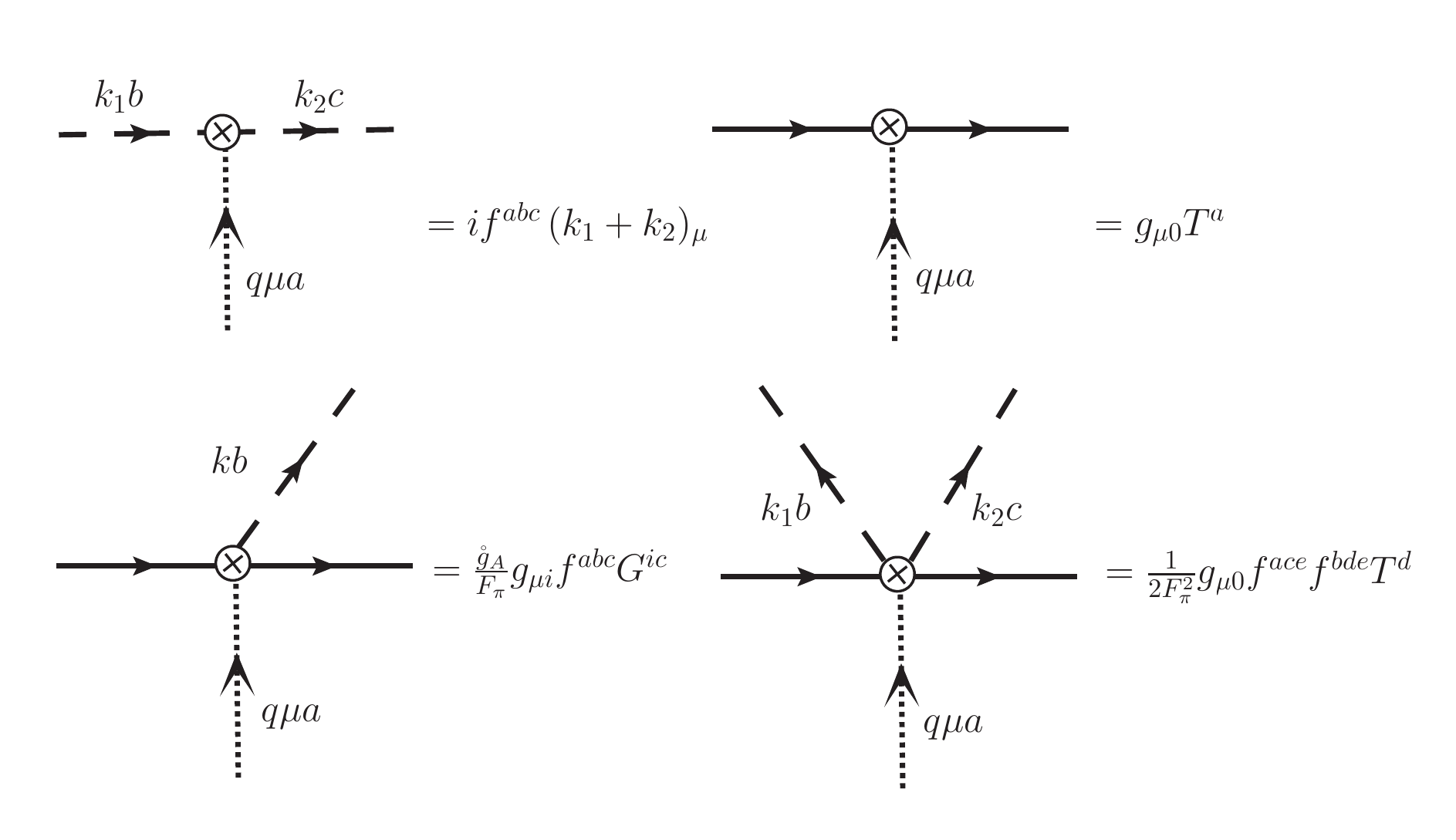}}
\caption{Vertices involving the vector currents from the LO Lagrangians.  }
\label{fig:vcurrents}
\end{figure}
\end{center}

\begin{center}
\begin{figure}[h]
\centerline{\includegraphics[width=10.cm,angle=-0]{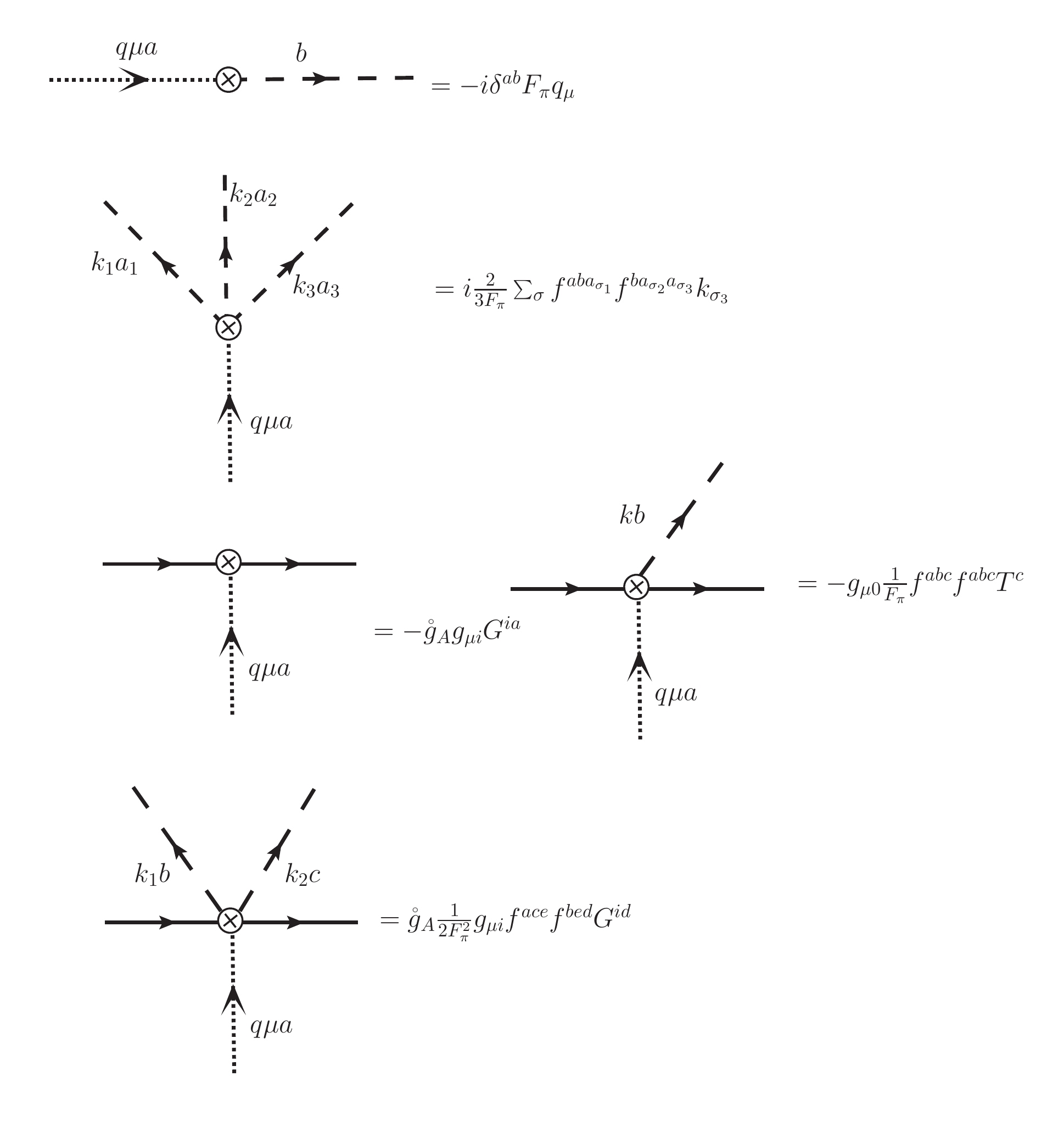}}
\caption{Vertices involving the axial-vector currents from the LO Lagrangians.  }
\label{fig:avcurrents}
\end{figure}
\end{center}

The interaction vertices and the currents derived from the LO Lagrangian and needed for the one-loop calculations are given here for convenience.
The  interactions are depicted in Fig.(\ref{fig:vertices}),  the vector    currents  in Fig.(\ref{fig:vcurrents}) and and the  axial-vector currents in Fig.(\ref{fig:avcurrents}).

\newpage

\section{Loop integrals}
\label{app:loopintegrals}
The one-loop integrals needed in this work are provided here.  The definition  $\widetilde{d^dk}\equiv d^d k/(2\pi)^d$ is used.

The scalar and tensor one-loop integrals are:
\bea
I(n,\alpha,\Lambda)\equiv\int \widetilde{d^dk}\; \frac{k^{2n}}{(k^2-\Lambda^2)^\alpha}&=&i(-1)^{n-\alpha} \frac{1}{(4\pi)^{\frac {d}{2}}} \frac{\Gamma(n+\frac{d}{2})\Gamma(\alpha-n-\frac{d}{2})}{\Gamma(\frac{d}{2})\Gamma(\alpha)}\left(\Lambda^2\right)^{n-\alpha+\frac{d}{2}}\nonumber\\
I^{\mu_1,\cdots,\mu_{2n}}(\alpha,\Lambda)\equiv\int \widetilde{d^dk}\; \frac{k_{\mu_1}\cdots k_{\mu_{2n}}}{(k^2-\Lambda^2)^\alpha}&=&i(-1)^{n-\alpha} \frac{1}{(4\pi)^{\frac{d}{2}}}\frac{1}{4^n n!} \frac{ \Gamma(\alpha-n-\frac{d}{2})}{ \Gamma(\alpha)}\left(\Lambda^2\right)^{n-\alpha+\frac{d}{2}}\nonumber\\
&\times& \sum_{\sigma} g_{\mu_{\sigma_1}\mu_{\sigma_2}}\cdots g_{\mu_{\sigma_{2n-1}}\mu_{\sigma_{2n}}} \\
&=&\frac{1}{4^n n!}\frac{\Gamma(\frac d2)}{\Gamma(n+\frac d2)} I(n,\alpha,\Lambda) \sum_{\sigma} g_{\mu_{\sigma_1}\mu_{\sigma_2}}\cdots g_{\mu_{\sigma_{2n-1}}\mu_{\sigma_{2n}}}~~,\nonumber
\eea
where $\sigma$ are the permutations of $\{1,\cdots,2n\}$.

The Feynman parametrizations needed when heavy propagators are in the loop are as follows:
\bea
\frac{1}{A_1\cdots A_m B_1\cdots B_n}&=& 2^m \Gamma(m+n) \int_0^\infty d\lambda_1\cdots d\lambda_m\int_0^1 d\alpha_1\cdots d\alpha_n \delta(1-\alpha_1-\cdots-\alpha_n)\nonumber\\
&\times& \frac{1}{(2\lambda_1 A_1+\cdots+2\lambda_m A_m+\alpha_1 B_1+\cdots+\alpha_n B_n)^{m+n}},
\eea
where the $A_i$ are heavy particle static propagators denominators, and the $B_i$ are relativistic ones.

The integration over a Feynman parameter $\lambda$ is of the general form:
\beq
J(C_0,C_1,\lambda_0,d,\nu)\equiv \int_0^\infty (C_0+C_1(\lambda-\lambda_0)^2)^{-\nu+\frac{d}{2}} d\lambda,
\eeq
which satisfies the recurrence relation:
\bea
J(C_0,C_1,\lambda_0,d,\nu)&=&\frac{-\lambda_0  (C_0+C_1\lambda_0^2)^{1-\nu+\frac{d}{2}}+(3+d-2\nu) J(C_0,C_1,\lambda_0,d,\nu-1)}{(d-2\nu+2)C_0}\nonumber\\
J(C_0,C_1,\lambda_0,d,\nu)&=&C_0 \frac{d-\nu}{d-2\nu+1}J(C_0,C_1,\lambda_0,d,\nu+1)+\frac{\lambda_0}{d-2\nu+1}(C_0+C_1 \lambda_0^2)^{\frac{d}{2}-\nu}.
\eea
Integrals with factors of $\lambda$ in the numerator are obtained by using
\bea
J(C_0,C_1,\lambda_0,d,\nu,n=1)&\equiv& \int_0^\infty (\lambda-\lambda_0)^{n=1} (C_0+C_1(\lambda-\lambda_0)^2)^{-\nu+\frac{d}{2}}d\lambda\nonumber\\
&=& -\frac{1}{2\, C_1\,(\frac{d}{2} + 1 - \nu)} (C_0+C_1\lambda_0^2)^{\frac{d}{2} + 1 - \nu},
\eea
and the recurrence relations
\beq
J(C_0,C_1,\lambda_0,d,\nu,n)=\frac{1}{C_1}(J(C_0,C_1,\lambda_0,d,\nu-1,n-1)-C_0 J(C_0,C_1,\lambda_0,d,\nu,n-2)).
\eeq
For   convenience in some of the calculations for the currents, the following integral is defined:
\beq
\tilde{J}(C_0,C_1,\lambda_0,d,\nu,n)\equiv J(C_0,C_1,\lambda_0,d,\nu,n)+\lambda_0 J(C_0,C_1,\lambda_0,d,\nu)
\eeq
 
 For the calculations in this work the following  integrals are needed at $d=4-2\eps$:
 \bea
J(C_0,C_1,\lambda_0,d,3)&=&\frac{1}{\sqrt{C_0 C_1}}\left(\frac{\pi}{2}+\arctan( \lambda_0\sqrt{\frac{C_1}{C_0}})\right) \nonumber\\
J(C_0,C_1,\lambda_0,d,2)&=&\frac{1}{d-3}(\lambda_0(C_0+C_1\lambda_0^2)^{\frac{d}{2}-2}+(d-4)C_0 J(C_0,C_1,\lambda_0,d,3))\nonumber\\
J(C_0,C_1,\lambda_0,d,1)&=&\frac{1}{d-1}(\lambda_0(C_0+C_1\lambda_0^2)^{\frac{d}{2}-1}+(d-2)J(C_0,C_1,\lambda_0,d,2))
 \eea

\subsection*{Specific integrals}
Here a summary of relevant one-loop integrals for the calculations in this work is provided for the convenience of the reader.

1) Loop integrals involving only relativistic propagators
\bea
I(0,1,M)&=&-\frac{i}{(4\pi)^{\frac d2}}\Gamma(1-\frac d2) M^{d-2}\nonumber\\
I(0,2,M)&=&\frac{i}{(4\pi)^{\frac d2}}\Gamma(2-\frac d2) M^{d-4}\nonumber\\
I(1,1,M)&=&\frac{i}{(4\pi)^{\frac d2}}\,\frac d2 \,\Gamma(-\frac d2) M^{d}\nonumber\\
I(1,2,M)&=&-\frac{i}{(4\pi)^{\frac d2}}\,\frac d2 \,\Gamma(1-\frac d2) M^{d-2}\nonumber\\
K(q,M_a,M_b)&\equiv&\int \widetilde{d^dk} \frac{1}{(k^2-M_a^2+i\eps)((k+q)^2-M_b^2+i\eps)}=\int_0^1 d\alpha \; I(0,2,\Lambda(\alpha))\nonumber\\
K^\mu(q,M_a,M_b)&\equiv&\int \widetilde{d^dk} \frac{k^\mu}{(k^2-M_a^2+i\eps)((k+q)^2-M_b^2+i\eps)}=\int_0^1 d\alpha \; (\alpha-1)\, q^\mu\,I(0,2,\Lambda(\alpha))\nonumber\\
K^{\mu\nu}(q,M_a,M_b)&\equiv&\int \widetilde{d^dk} \frac{k^\mu k^\nu}{(k^2-M_a^2+i\eps)((k+q)^2-M_b^2+i\eps)}\nonumber\\
&=&\int_0^1 d\alpha \;( (1-\alpha)^2\, q^\mu q^\nu\,I(0,2,\Lambda(\alpha))+\frac{g^{\mu\nu}}{d} I(1,2,\Lambda(\alpha))),
\eea
where:
\beq
\Lambda(\alpha)=\sqrt{\alpha M_a^2+(1-\alpha)M_b^2-\alpha(1-\alpha) q^2}\nonumber
\eeq
2) Loop integrals involving one heavy propagator
\bea
H(p^0,M)&\equiv&\int \widetilde{d^dk} \frac{1}{(p^0-k^0+i\eps)(k^2-M^2+i\eps)}\nonumber\\
&=&\frac{2i}{(4\pi)^{\frac d2}}\Gamma(2-\frac d2) J(M^2-p^{0^{\mbox{\scriptsize  2}}},1,p^0,d,2)   \nonumber\\
H^{ij}(p^0,M)&\equiv& \int \widetilde{d^dk} \frac{k^i k^j}{(p^0-k^0+i\eps)(k^2-M^2+i\eps)}\nonumber\\
&=& -\frac{i}{(4\pi)^{\frac d2}} g^{ij}\Gamma(1-\frac d2)J(M^2-p^{0^{\mbox{\scriptsize  2}}},1,p^0,d,1)
 \\
  H^{ij\mu}(p^0,M_a,M_b,q)&\equiv&\int \widetilde{d^dk} \frac{k^i (k+q)^j (2k+q)^\mu}{(p^0-k^0+i\eps)(k^2-M_a^2+i\eps)((k+q)^2-M_b^2+i\eps)}\nonumber\\
  &=&i\frac{4}{(4\pi)^{\frac d2}}\int_0^1 d\alpha\left\{ -\frac 12 \Gamma(3-\frac d2) q^iq^j \alpha(1-\alpha)\right.\nonumber\\
  &\times& \left((1-2\alpha)q^\mu J(C_0,C_1,\lambda_0,d,3)-2\,g^{\mu 0}  \tilde{J}(C_0,C_1,\lambda_0,d,3,1)\right)\nonumber\\
  &+& \Gamma(2-\frac d2)\left((-(1-2\alpha)g^{ij} q^\mu+2(\alpha g^{\mu i} q^j-(1-\alpha) g^{\mu j} q^i)) J(C_0,C_1,\lambda_0,d,2)\right.\nonumber\\
&+&  \left.\left. 2 g^{ij} g^{\mu 0} \tilde{J}(C_0,C_1,\lambda_0,d,2,1)\right)\right\},\nonumber
\eea
where:
\bea
C_0&=& \alpha M_a^2+(1-\alpha) M_b^2-p^{0^{\mbox{\scriptsize  2}}}-2(1-\alpha) p^0 q^0-(1-\alpha)(\alpha\, q^2+(1-\alpha)q^{0^{\mbox{\scriptsize  2}}})\nonumber\\
C_1&=& 1\nonumber\\
\lambda_0&=& p^0+(1-\alpha) q^0.
\eea
The polynomial pieces of the integrals are as follows:
\bea
H(p^0,M)^{\text{poly}}&=&\frac{i}{(4\pi)^2} 2 p^0(\lambda_\eps+2)\nonumber\\
H^{ij}(p^0,M)^{\text{poly}}&=&\frac{i}{(4\pi)^2} \frac{p^0}{3}((3M^2-2p^{0^{\mbox{\scriptsize  2}}})\lambda_\eps+7 M^2-\frac{16}{3}p^{0^{\mbox{\scriptsize  2}}})\nonumber\\
H^{ij0}(p^0,M_a,M_b,q)^{\text{poly}}&=&\frac{i}{6(4\pi)^2}\left( (2 q^iq^j+q^2 g^{ij})\lambda_\eps+q^2 g^{ij}-3(\lambda_\eps+1)(M_a^2+M_b^2) g^{ij}\right.\nonumber\\
&+&\left. 3(\lambda_\eps+2)(2p^0+q^0)^2 g^{ij}\right),
\eea
where the UV divergency is given by the terms proportional to $\lambda_\eps\equiv 1/\eps-\gamma+\log 4\pi$, where $d=4-2\eps$.

 \section{Useful operator reductions}
 \label{app:opred}
 
The reductions of multi-body spin-flavor operators which appear in the polynomial contributions of  the  one-loop corrections  to the self-energy and the currents  require some lengthy work, and are therefore  provided here. The reductions are only valid for matrix elements between states in the totally symmetric irreducible representation of $SU(6)$. In the following $\delta \hat m$ contains only the hyperfine term.
 
1)  Self-energy:
 \bea
 \!\,[[\delta \hat{m},G^{ia}],G^{ia}]&=&\frac{C_{\rm HF}}{N_c}\left(\frac{7}{2}\hat S^2-\frac{3}{8}N_c(N_c+6)\right)\nonumber\\
 	 \!\,[[\delta \hat{m},[\delta \hat{m},G^{ia}]],G^{ia}]&=&\left(\frac{C_{\rm HF}}{N_c}\right)^2
	 \left(4\hat S^4-(N_c(N_c+6)-18)\hat S^2-\frac{3}{2} N_c(N_c+6)\right)\nonumber\\
  \!\,[[\delta \hat{m},[\delta \hat{m},[\delta \hat{m},G^{ia}]]],G^{ia}]&=&	 
  \left(\frac{C_{\rm HF}}{N_c}\right)^3\left(36 \hat S^4-(5N_c(N_c+6)-36)\hat S^2-3 N_c(N_c+6)\right)\nonumber\\
   M_a^2 G^{ia} G^{ia}&=& 2B_0\left(m^0 \hat G^2+m^a(-\frac{7}{24}\{S^i,G^{ia}\}+\frac{3}{16}(N_c+3) T^a)\right)\nonumber\\
 M_a^2 [[\delta \hat{m},G^{ia}],G^{ia}]&=& 4\frac{C_{\rm HF}}{N_c} B_0 \left(\frac{8}{3}m^0 \hat S^2+\frac{5}{12}m^a \{S^i,G^{ia}\}\right)-4 M_a^2 G^{ia} G^{ia}
  \label{OpRedMasses}
 \eea
 
 2) Vector currents:
 \bea
 G^{ia}[\delta \hat{m},[\delta \hat{m},G^{ia}]]&=&\left(\frac{C_{\rm HF}}{N_c}\right)^2\left(\frac 34N_c(N_c+6)+(\frac 12 Nc(N_c+6)-9)\hat S^2-2 \hat S^4\right)\nonumber\\
 ~[\delta \hat{m},G^{ia}] [\delta \hat{m}, G^{ia}]&=&  -G^{ia}[\delta \hat{m},[\delta \hat{m},G^{ia}]]\nonumber\\
  G^{ib} T^a [\delta \hat{m},[\delta \hat{m},G^{ib}]]&=&-[\delta \hat{m},G^{ib}]\;T^a\;[\delta \hat{m},G^{ib}]\nonumber\\
  &=& \left(\frac{C_{\rm HF}}{N_c}\right)^2\bigg (3(N_c+3)S^i G^{ia} \nonumber\\ 
  &+&\left.\left(\frac 34(N_c(N_c+6)-6)+\frac 12(N_c(N_c+6)-30)\hat S^2-2\hat S^4\right)T^a\right)\nonumber\\
 \!\,[[T^a,G^{ib}],[\delta \hat{m},[\delta \hat{m},G^{ib}]]]&=& -[[T^a, [\delta \hat{m},G^{ib}]],[\delta \hat{m},G^{ib}]]\nonumber\\
 &=&2 [\delta \hat{m},G^{ib}]T^a  [\delta \hat{m},G^{ib}]-\{T^a, [\delta \hat{m},G^{ib}] [\delta \hat{m},G^{ib}]\}\nonumber\\
  f^{abc}f^{bcd} M_b^2 T^d&=& 6 B_0 \left(m^0 T^a+\frac{1}{4} d^{abc} m^b T^c\right)\nonumber\\
 M_b^2 G^{ib} T^a G^{ib}&=&2B_0\left (m^0(\hat G^2-\frac 98)T^a+\frac 12 m^b\left (\frac 12 \{T^a,\frac 38 (N_c+3) T^b-\frac{7}{24} S^i G^{ib}\} -\frac 34 d^{abc}T^c\right)\right)\nonumber\\
 M_b^2[[T^a,G^{ib}],G^{ib}]&=&\frac{9}{2} B_0\left(  m^0 T^a+\frac 14 m^b d^{abc}T^c\right)
 \eea

 3) Axial-vector currents:
 
 \bea
 G^{jb}G^{ia}[\delta \hat{m},[\delta \hat{m},G^{jb}]]+h.c.&=&\left(\frac{C_{\rm HF}}{N_c}\right)^2\left(\frac 32 N_c(N_c+6) G^{ia}+\left(\frac 12 N_c(N_c+6)-14\right)\{\hat S^2,G^{ia}\}\right.\nonumber\\
 &-&\left. \{\hat S^2,\{\hat S^2,G^{ia}\}\}+\frac 32(N_c+3) S^i T^a+2S^i S^j G^{ja}\right)\nonumber\\
 ~[\delta \hat{m},G^{jb}]G^{ia}[  \delta \hat m,G^{jb}]&=&\left(\frac{C_{\rm HF}}{N_c}\right)^2\left(-\frac 12\left(3+\frac 12N_c(N_c+6)\right) G^{ia}\right.\nonumber\\
&+&\left.\frac 12\left( 13-\frac 12N_c(N_c+6)\right)\{\hat S^2,G^{ia}\} + \frac 12\{\hat S^2,\{\hat S^2,G^{ia}\}\}-\frac 54(N_c+3)S^iT^a\right)\nonumber\\
 f^{acd}f^{bcd} M_c^2G^{ib}&=& 6B_0\left(m^0 \delta^{ab}+\frac 14 m^c d^{abc}\right)G^{ib}\nonumber\\
 M_b^2 G^{jb}G^{ia}G^{jb}&=&\frac 12\{G^{ia},M_b^2 G^{jb} G^{jb}\}- \frac{B_0}{12}\left(23 \,m^0 G^{ia}  +   m^b\left(\frac 53 \delta^{ab} S^i+\frac{11}{4} d^{abc} G^{ic}\right)\right)
\eea

\section{Figures for the fits to LQCD and physical masses}
 \label{app:LQCDmasses}
 \begin{figure}[h]
\centering
\scalebox{.71}{
\begin{tabular}{ccc}
\includegraphics[width=.45\linewidth]{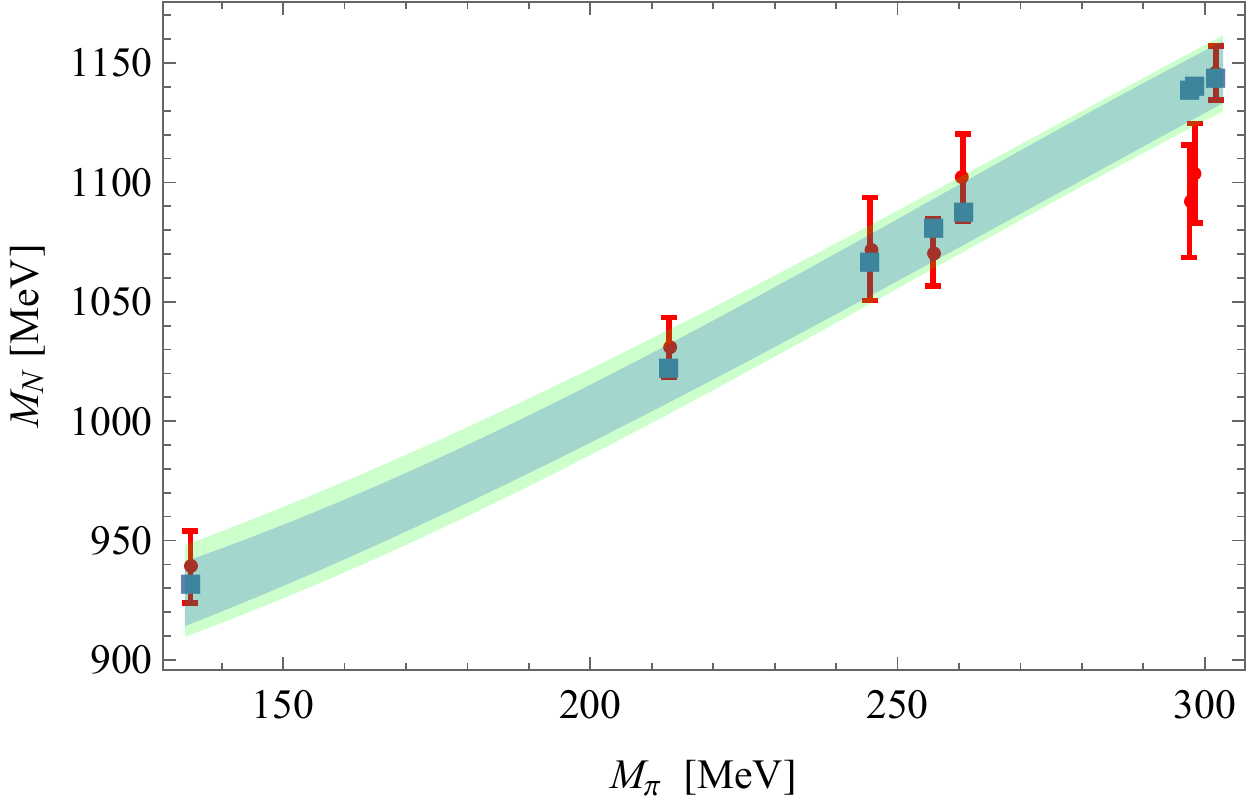} & \includegraphics[width=.45\linewidth]{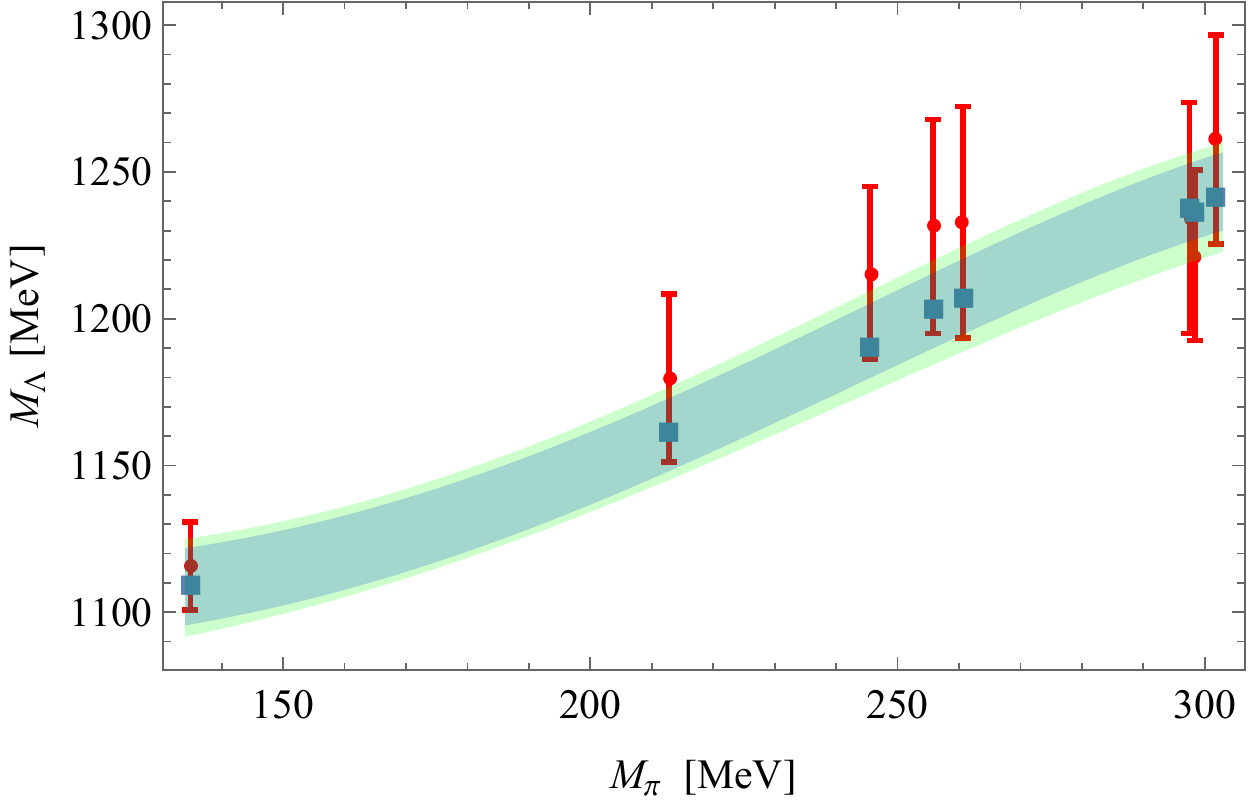}&
\includegraphics[width=.45\linewidth]{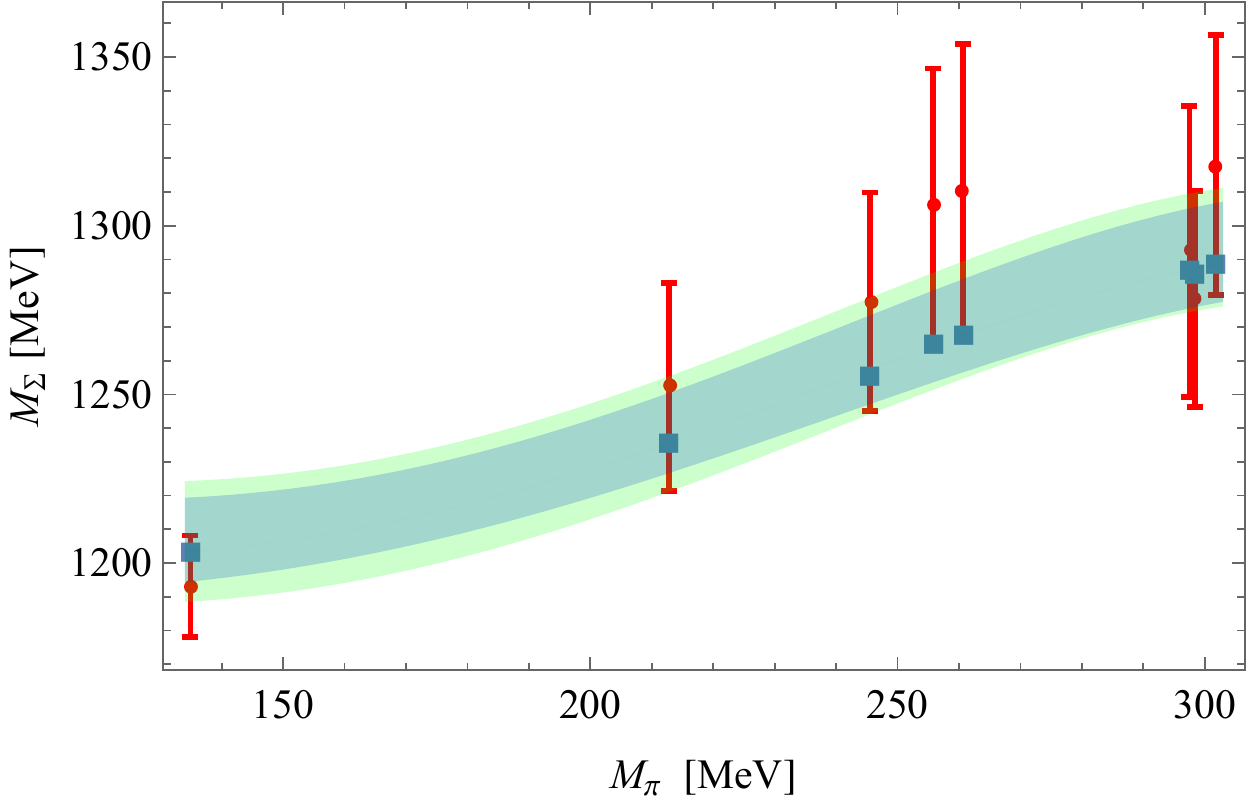}\\
\includegraphics[width=.45\linewidth]{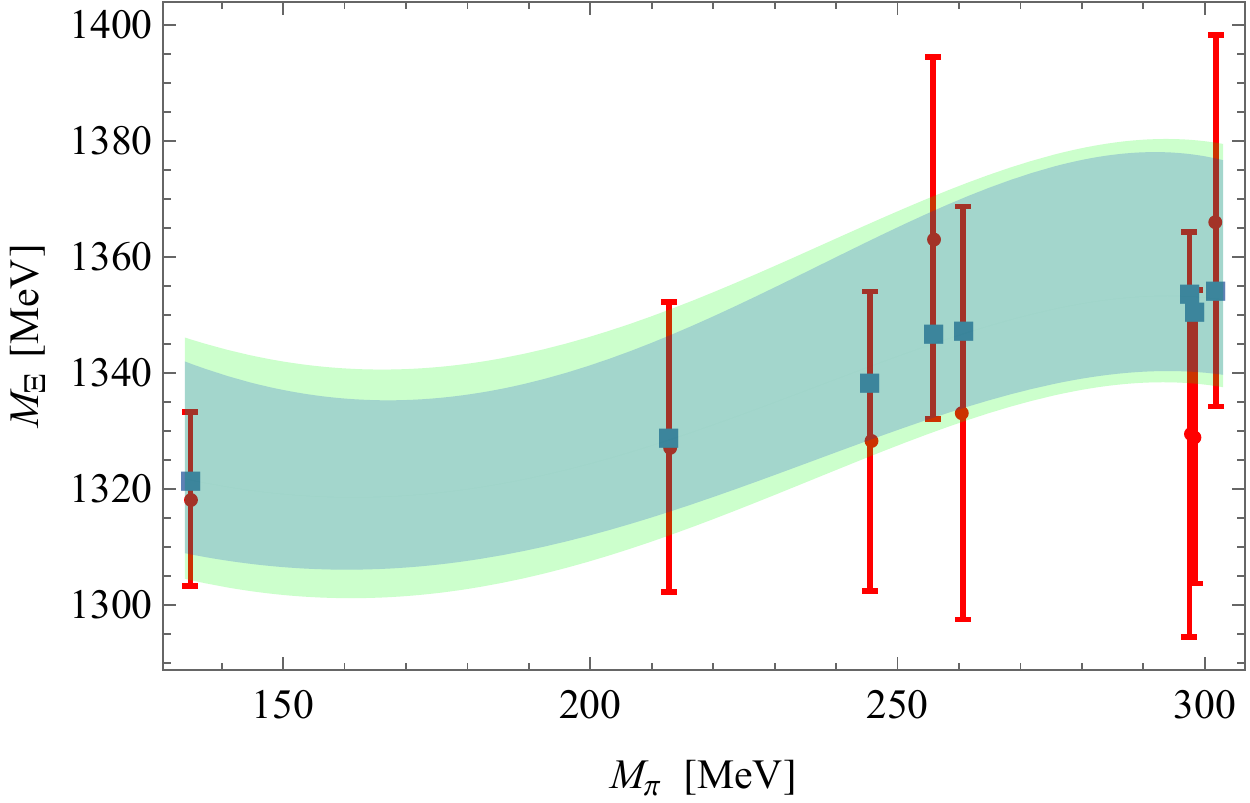}&    \includegraphics[width=.45\linewidth]{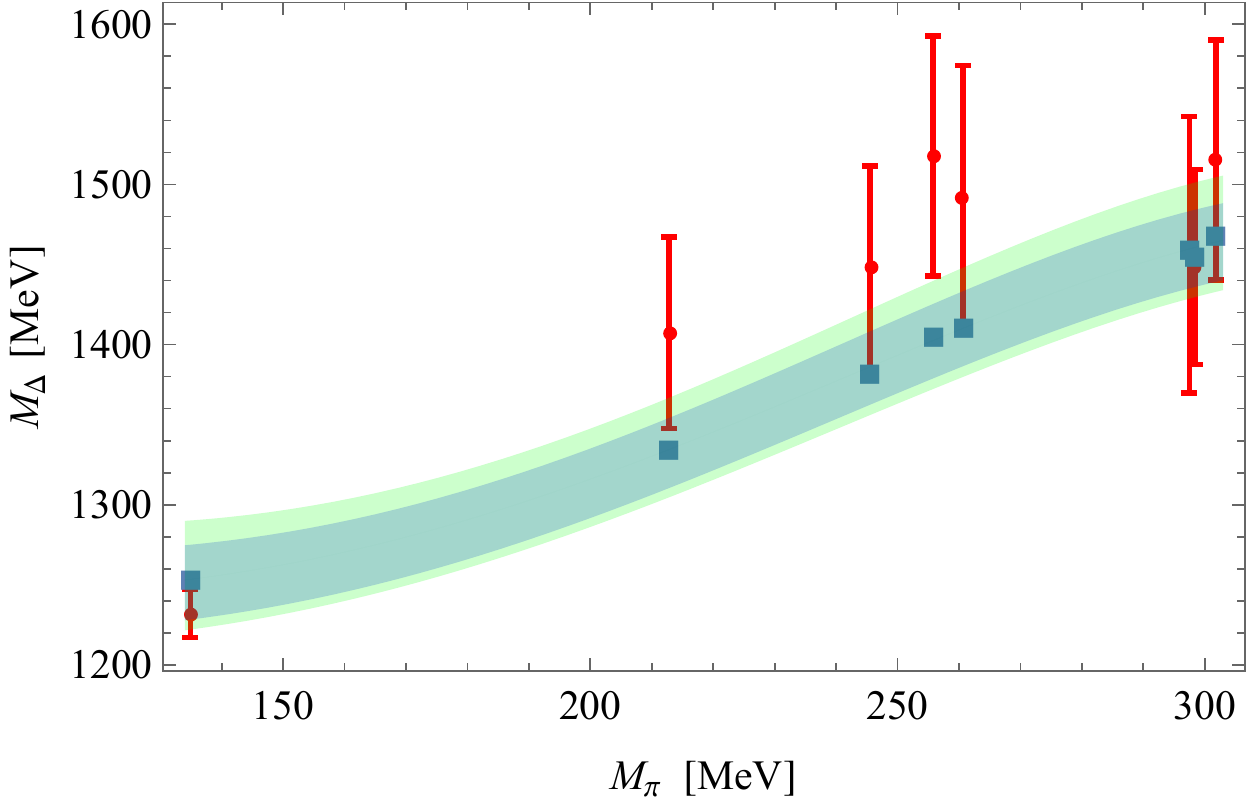} & \includegraphics[width=.45\linewidth]{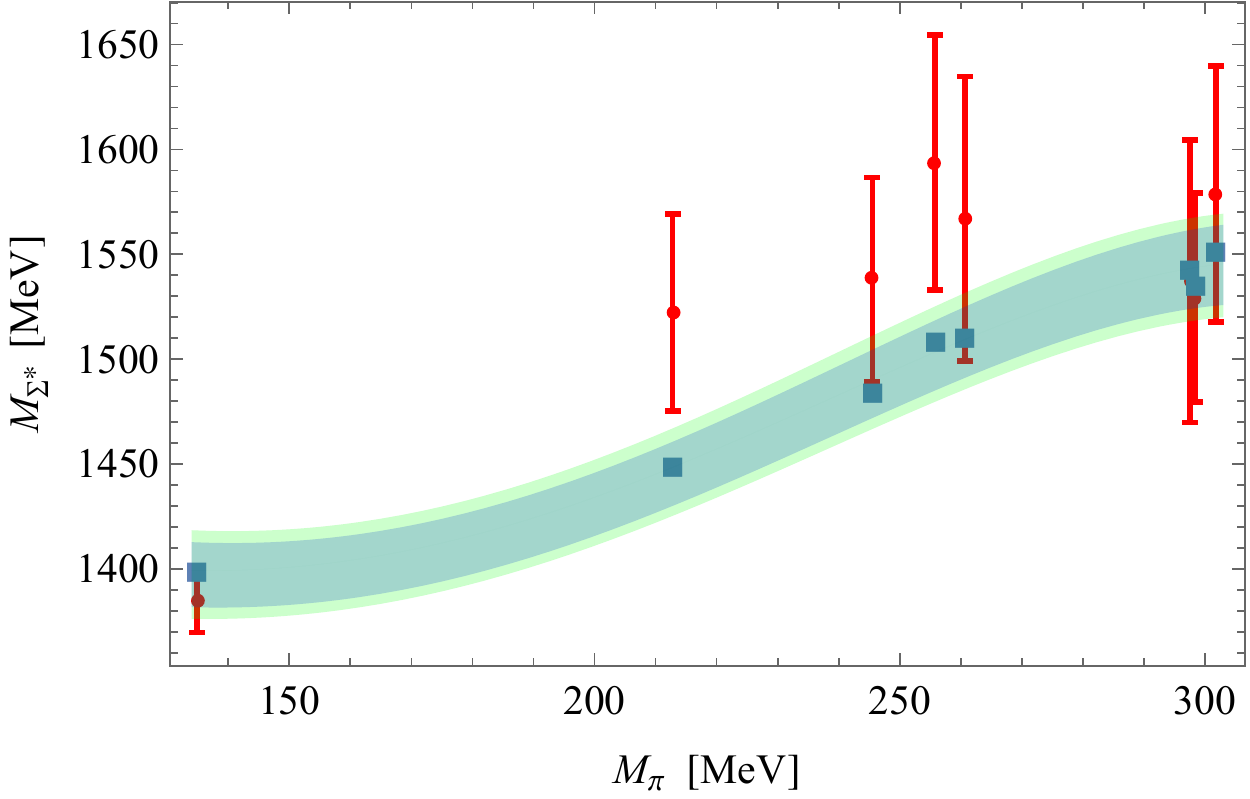}\\
\end{tabular}}
\scalebox{.71}{\begin{tabular}{cc}
\includegraphics[width=.45\linewidth]{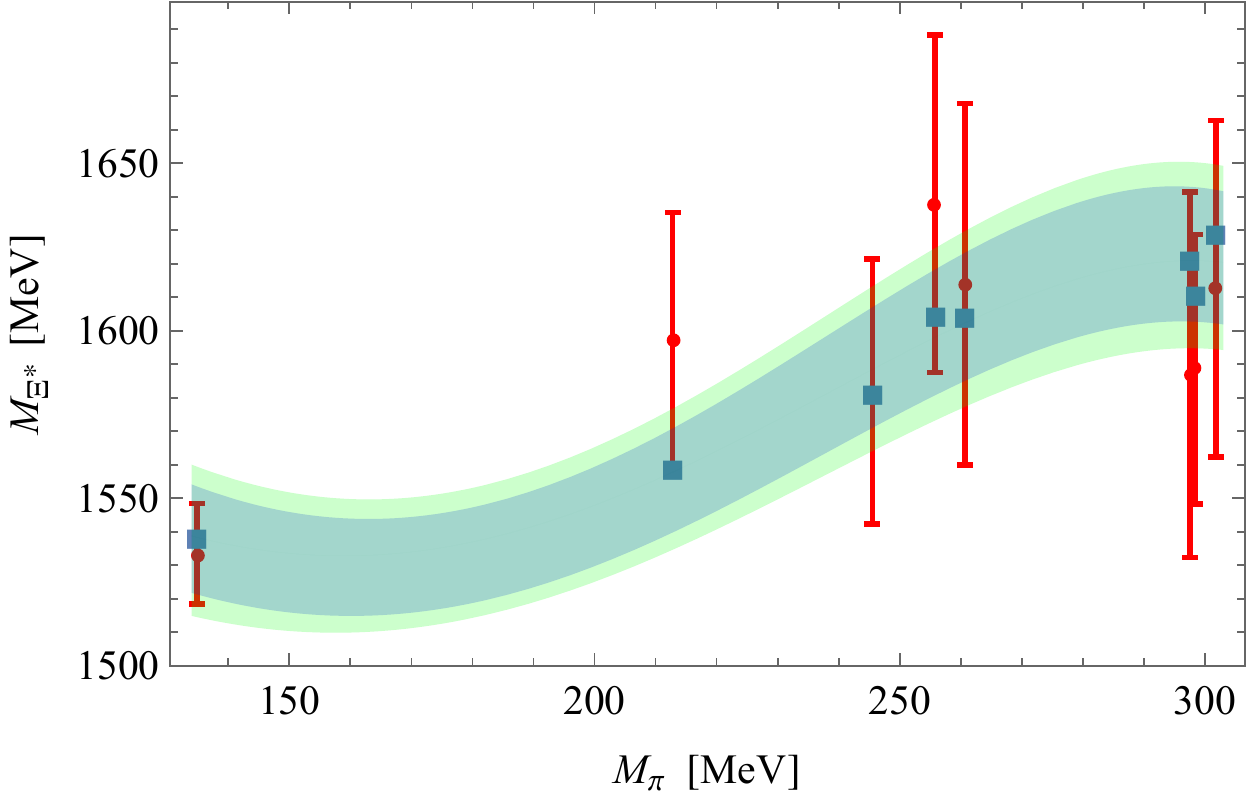} & \includegraphics[width=.45\linewidth]{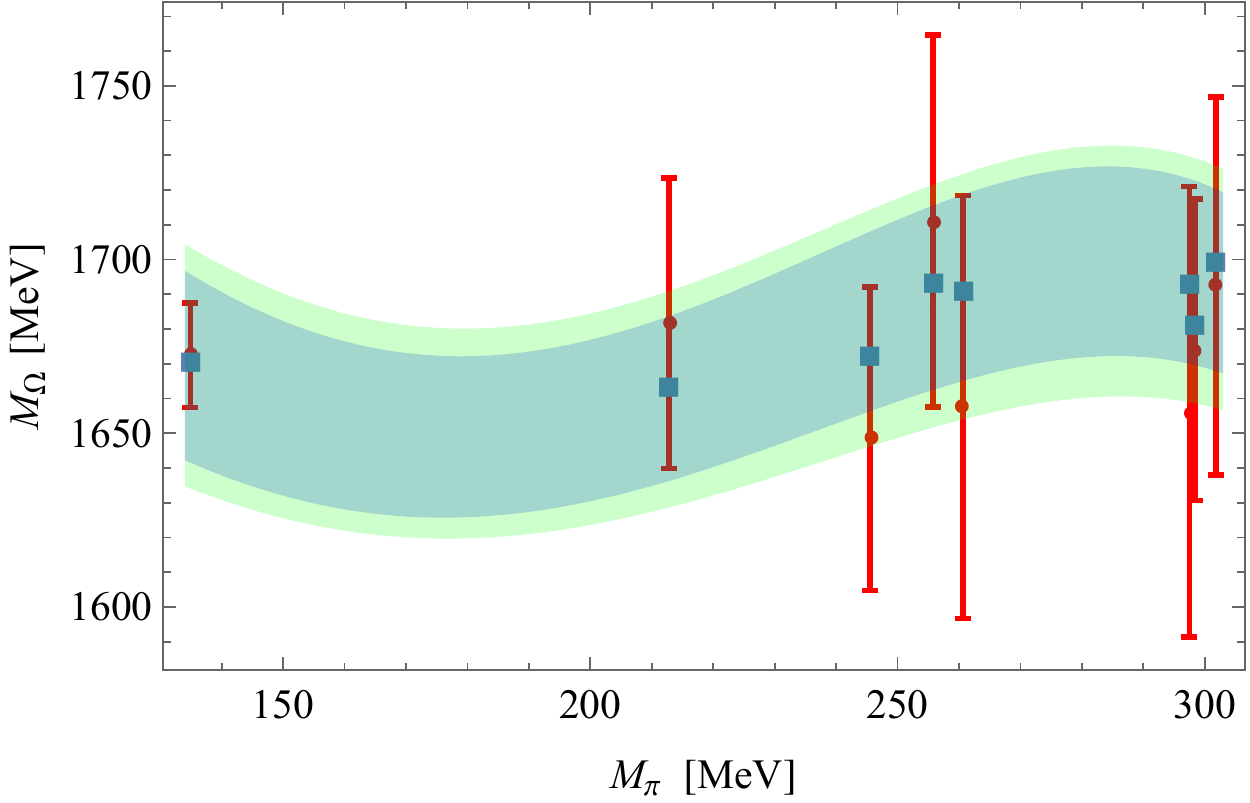}  
\end{tabular} }
\caption{Baryon masses vs $M_\pi$ obtained from the combined fit (second row of Table\thinspace(\ref{tab:massfits})). The bands correspond to the 67\%  and 95\% confidence intervals.   The red points with error bars are  from the LQCD calculations \cite{Alexandrou:2014sha},  and the squares are the theoretical values for the values of $M_\pi$ and $M_K$ of the corresponding data point. }
\label{fig:BChPT-combined-fits}
\end{figure}

\newpage
 
\bibliography{Refs}

\end{document}